\documentclass[final,1p,times]{elsarticle}

\usepackage{graphicx}
\usepackage{booktabs} % For formal tables
\usepackage[T1]{fontenc}
\usepackage[utf8]{inputenc}
\usepackage{hyperref}
\usepackage{amsmath} % theoremstyls
\usepackage{listings}
\usepackage[textsize=scriptsize]{todonotes} 
\usepackage{fancybox}
\usepackage{wrapfig}
\graphicspath{{figures/}}
\usepackage{lscape}
\usepackage{longtable}
\usepackage{tabularx}
\usepackage{adjustbox}
\usepackage[linesnumbered,ruled,vlined]{algorithm2e}
\usepackage[T1]{fontenc}

\newcommand{\myurl}[1]{{\fontsize{7}{8}\selectfont{\url{#1}}}}
\newcommand{\myequationfont}{
\fontsize{8}{9}\selectfont
}
\newcommand{\mycodefont}{
\fontsize{8}{9}\selectfont\ttfamily
}

\newcommand{\code}[1]{\texttt{#1}}
\newcommand{\eg}{e.g.,~}

\newcommand{\qq}[1]{``#1''}

\journal{Revised version}

\begin{document}

\begin{frontmatter}


\title{Template-Based Question Answering over Linked Geospatial Data}



\author[mymainaddress]{Dharmen Punjani}
\author[mymainaddress]{Markos Iliakis}
\author[mymainaddress]{Theodoros Stefou}
\author[mysecondaryaddress]{Kuldeep Singh}
\author[mysecondaryaddress2,mysecondaryaddress3]{Andreas Both}
\author[mymainaddress]{Manolis Koubarakis}
\author[mymainaddress]{Iosif Angelidis}
\author[mymainaddress]{Konstantina Bereta}
\author[mymainaddress]{Themis Beris}
\author[mymainaddress]{Dimitris Bilidas}
\author[mymainaddress]{Theofilos Ioannidis}
\author[mymainaddress]{Nikolaos Karalis}
\author[mysecondaryaddress4]{Christoph Lange}
\author[mymainaddress]{Despina-Athanasia Pantazi}
\author[mymainaddress]{Christos Papaloukas}
\author[mymainaddress]{Georgios Stamoulis}
%\author[affiliation]{Dharmen Punjani}
%\affiliation{Dept. of Informatics and Telecommunications, National and Kapodistrian University of Athens}

\address[mymainaddress]{Department of Informatics and Telecommunications, National and Kapodistrian University of Athens\linebreak \{dpunjani,koubarak, iosang, konstantina.bereta, tberis, d.bilidas, tioannid, nkaralis, langec, dpantazi, christospap, gstam\}@di.uoa.gr}
\address[mysecondaryaddress]{Cerence GmbH, Aachen, Germany, kuldeep.singh1@cerence.com}
\address[mysecondaryaddress2]{DATEV eG, Germany, contact@andreasboth.de}
\address[mysecondaryaddress3]{Anhalt University of Applied Science, Köthen, Germany, andreas.both@hs-anhalt.de}
\address[mysecondaryaddress4]{Fraunhofer FIT, Sankt Augustin, Germany, lange@informatik.rwth-aachen.de}

%\author{ Dharmen Punjani$^1$, Markos Iliakis$^1$, Kuldeep Singh$^2$, Andreas Both$^3$, Manolis Koubarakis$^1$, Iosif Angelidis$^1$, Konstantina Bereta$^1$, Themis Beris$^1$, Dimitris Bilidas$^1$, Theofilos Ioannidis$^1$, Nikolaos Karalis$^1$,Christoph Lange$^2$, Despina-Athanasia Pantazi$^1$, Christos Papaloukas$^1$, Georgios Stamoulis$^1$}
%\authornote{Short name for the following authors: I. Angelidis, K. Bereta, T. Beris, D. Bilidas, A. Both, T. Ioannidis, N. Karalis,  M. Koubarakis, D. Pantazi, C. Papaloukas, D. Punjani and G. Stamoulis.}
%\affiliation{
 % \small \institution{$^1$ Dept. of Informatics and Telecommunications, National and Kapodistrian University of Athens \linebreak \{dpunjani, koubarak\}@di.uoa.gr,
 % \linebreak $^2$ Fraunhofer IAIS and University of Bonn \linebreak kuldeep.singh@iais.fraunhofer.de, langec@cs.uni-bonn.de,
%\linebreak $^3$ DATEV eG, Germany \linebreak contact@andreasboth.de }
%  \streetaddress{University Campus, Ilisia}
%  \city{Athens}
% \state{Greece}
%  \postcode{15784}
%}
\begin{abstract}
Large amounts of geospatial data have been made available recently on the linked open data cloud and the portals of many national cartographic agencies (e.g., OpenStreetMap data, administrative geographies of various countries, or land cover/land use data sets). These datasets use various geospatial vocabularies and can be queried using SPARQL or its OGC-standardized extension GeoSPARQL. In this paper, we go beyond these approaches to offer a question-answering engine for natural language questions on top of linked geospatial data sources. Our system has been implemented as re-usable components of the Frankenstein question answering architecture. We give a detailed description of the system's architecture, its underlying algorithms, and its evaluation using a set of 201 natural language questions. The set of questions is offered to the research community as a gold standard dataset for the comparative evaluation of future geospatial question answering engines.
\end{abstract}

\begin{keyword}
Linked Geospatial Data \sep General Administrative Divisions dataset (GADM) \sep OpenStreetMap \sep Question Answering \sep GeoSPARQL \sep Information Retrieval \sep Semantic web


\end{keyword}

\end{frontmatter}


\section{Introduction}\label{sec:introduction}
%%%%%%%%%%%%%%%%%%%%%%%%%%%%%%%%%%%%%%%%%%%%%%%%%%%%%%%%%%%%%%%%%%%%%%%%%%%%%%%%%%%%%%%%%%%%%
%%%%%%%%%%%%%%%%%%%%%%%%%%%%%%%%%%%%%%%%%%%%%%%%%%%%%%%%%%%%%%%%%%%%%%%%%%%%%%%%%%%%%%%%%%%%%
The number of data sources in private environments, enterprises, and the Web is increasing continuously, increasing the effort of making data accessible. One important means of making data accessible is \emph{question answering (QA)}, which provides a natural language interface for common users to express their information needs \cite{DBLP:journals/nle/HirschmanG01}. Users commonly pose questions or information requests with a \emph{geospatial dimension} to search engines, e.g., \qq{Christmas market in Germany}, \qq{Schools in London}, \qq{Is there a Macy's near Ohio?}, \qq{Which countries border Greece?}.
%These are typical trending questions of real users as shown on Google Trends\footnote{\url{https://trends.google.de/trends/}}.
Answering such questions or information requests requires data that has a geospatial dimension as well.

\emph{Geospatial} or \emph{geographic knowledge} has been studied for many years by researchers in Geography, Geographic Information Systems (GIS), Geographic Information Retrieval (GIR), Databases, Artificial Intelligence and the Semantic Web, and there is a wealth of research results  concerning representation, querying and inference for geographic knowledge. In GIS terminology which we use in this paper, a
\emph{geographic feature} (or simply \emph{feature}) is an abstraction of a real world phenomenon and can have various attributes that describe its \emph{thematic} and \emph{spatial} characteristics. For example, the country Greece is a feature, its name and population are thematic attributes, while its location on Earth in terms of polar co-ordinates is a spatial attribute. Knowledge about the spatial attributes of a feature can be \emph{quantitative} or \emph{qualitative}. For example, the fact that the distance between Athens and Salonika is 502 km is quantitative knowledge, while the fact that river Evros crosses Bulgaria and Turkey and is at the border of Greece with Turkey is qualitative knowledge.
Quantitative geographic knowledge is usually represented using \emph{geometries} (e.g., points, lines and polygons on the Cartesian plane) while qualitative geographic knowledge is captured by \emph{qualitative binary relations} between the geometries of features.

A significant fraction of the available data on the Web is geospatial, and this fraction is growing by 20 percent or more per year \cite{DBLP:journals/bdr/LeeK15}.
Qualitative Geospatial data can be expressed as a property of an entity or an explicit assertion. For example, in the RDF dataset DBpedia\footnote{\url{http://wiki.dbpedia.org/}} extracted from Wikipedia, the
resource {\tt dbr:Berlin} has a data property {\tt dbo:Country} with value {\tt dbr:Germany} enabling the answering of questions such as \qq{Cities in Germany} using DBpedia. Or a dataset like DBpedia can contain the fact {\tt dbr:Berlin ogc:sfWithin dbr:Germany} and the question \qq{Which cities are in Germany?}
can again be answered using this dataset.
Quantitative Geospatial data
%(\ie not represented by an explicit data property)
can be expressed by a property of an entity which has value a geometry (latitude/longitude pair or polygon).
Then, the question \qq{Which cities are within 100 km of Berlin?}\ can be answered by retrieving the geometry of the resource {\tt dbr:Berlin} from an appropriate geospatial dataset, and then computing the distance of
this geometry to the geometries of cities outside Berlin.
In this paper, we focus on question answering from qualitative and quantitative geospatial knowledge made available on the Web as \emph{linked open data.}

Examples of geospatial data published on the Web as linked open data include geospatial data from various countries (e.g., the United Kingdom\footnote{\url{http://data.ordnancesurvey.co.uk/}} or The Netherlands\footnote{\url{https://www.kadaster.nl/-/bag-linked-data}}), OpenStreetMap data (published in RDF by project LinkedGeoData\footnote{\url{http://linkedgeodata.org/About}}~\cite{DBLP:conf/semweb/AuerLH09} but also by our AI group at the National and Kapodistrian University of Athens\footnote{\url{http://ai.di.uoa.gr/\#datasets}}), and land cover/land use data sets (e.g., the European CORINE land cover dataset published in RDF by our group in the context of various European projects).
%These datasets use various vocabularies and can be queried by pure SPARQL, the Open Geospatial Consortium (OGC) standard GeoSPARQL or stSPARQL, a  query language that supports time as well as space~\cite{strabon}.
Queries over such data can be asked using the linked data query language  SPARQL and its geospatial extensions GeoSPARQL\footnote{\url{http://www.opengeospatial.org/standards/geosparql}} and stSPARQL~\cite{DBLP:conf/semweb/KyzirakosKK12}.
%, and use a system such as Strabon~\cite{strabon}, Ontop-spatial~\cite{ontop-spatial} or GraphDB\footnote{\url{http://ontotext.com/products/graphdb/}} to store the relevant data.
However, to better serve the needs of non-technical end users, it would be worthwhile to offer a natural language interface to linked geospatial data based on QA techniques. To the best of our knowledge, none of the QA systems utilizing linked data developed in recent years 
%(see 
%\cite{DBLP:journals/semweb/HoffnerWMULN17} and 
\cite{DBLP:journals/kais/DiefenbachLSM18} 
%for a recent overview) 
deals with geospatial data. The work presented in this paper makes the first steps towards this direction.
%this paper.

%\emph{geospatial} or \emph{geographic} information, although such knowledge is extremely important, given that many pieces of information inherently have some spatial dimension.

%, meanwhile a well-established way of satisfying information needs intuitively thanks to natural language input,

In a similar spirit, the GIR community has been emphasizing the need to develop techniques for answering geographic questions expressed in natural language over text data since 2004.\footnote{\url{http://www.geo.uzh.ch/~rsp/gir18/}}
The importance of the research issues studied in GIR can also be seen by the fact that interaction with geospatial search engines on mobile devices today is often done using spoken natural language (\eg in Google Maps you can ask for directions to a place and these directions are then spoken to you).
This is in agreement with the vision of multimodal spatial querying presented in~\cite{DBLP:conf/hci/SchlaisichE01}. Geographical knowledge is also very important in the new generation of personal assistants such as Amazon Alexa or Google Home.

Important assets of the GIR and QA research communities are the \emph{gold standards} i.e., datasets and sets of questions that can be used to test the effectiveness of developed systems and perform detailed comparisons of them. In the area of QA over linked data, such gold standards have recently been provided by the QALD challenge\footnote{\url{https://qald.sebastianwalter.org/}}.
To the best of our knowledge, no gold standard for geospatial question answering  over linked data has been proposed so far by QALD or any other relevant research activity.

The contributions of this paper are the following. We have designed and implemented GeoQA, the {\it first} question answering system for linked geospatial data. GeoQA is implemented using reusable components as part of the component-oriented Qanary question answering methodology~\cite{DBLP:conf/esws/SinghBDSC016,DBLP:conf/esws/BothDSSC016} and its most recent implementation Frankestein~\cite{DBLP:conf/www/SinghRBSLUVKP0V18}.

We have also developed a gold standard for question answering over linked geospatial data which consists of two parts. The first part is a linked geospatial dataset built from DBpedia, the GADM database of global administrative areas\footnote{\url{http://www.gadm.org/}} and OpenStreetMap (OSM)\footnote{\url{https://www.openstreetmap.org}}. For the purposes of the gold standard, GADM and OSM have been restricted to the United Kingdom and Ireland. The second part of the gold standard consists of 201 geospatial questions that have been collected by student volunteers at the National and Kapodistrian University of Athens. The gold standard is used in a  evaluation of the effectiveness of GeoQA and it is also made freely available to the research community for evaluating other future proposals.\footnote{\url{http://geoqa.di.uoa.gr/}}
In this way, we contribute to a long-term research agenda towards question answering systems with geospatial features.

A previous version of this paper has been presented in the
12th Workshop on Geographic Information Retrieval (GIR'18)~\cite{punjani2018template}.
The current version of the paper contains the following additional contributions:
\begin{itemize}
\item We added
support to the question patterns of Categories 1 and 6, and
count questions of Category 7 of the gold standard question set.
\item We have performed a detailed study of existing named entity recognizers and disambiguators
to select the one we have used in the GeoQA engine (see Section 5.2).
\item We performed various optimizations and code improvements in the instance identifier, property identifier
and query generator of our engine (see Section 5).
\item As a result of the above, we have now 
achieved much higher precision, recall and f-measures compared to the
previous version of the GeoQA engine presented in~\cite{punjani2018template}.
\end{itemize}

The rest of the paper is organized as follows.
The next section presents related work.
Section \ref{sec:GeoData} presents the three datasets of the gold standard and the interlinking of GADM and OSM with DBpedia.
Section \ref{sec:Benchmark} presents the gold standard questions.
%The next section presents the Qanary framework.
Section \ref{sec:geopatial} presents our approach to building the query answering pipeline used by GeoQA.
Section \ref{sec:evaluation} presents an evaluation of GeoQA using the gold standard.
Section \ref{sec:conclusion} concludes the paper and discusses future work.

%%%%%%%%%%%%%%%%%%%%%%%%%%%%%%%%%%%%%%%%%%%%%%%%%%%%%%%%%%%%%%%%%%%%%%%%%%%%%%%%%%%%%%%%%%%%%
%%%%%%%%%%%%%%%%%%%%%%%%%%%%%%%%%%%%%%%%%%%%%%%%%%%%%%%%%%%%%%%%%%%%%%%%%%%%%%%%%%%%%%%%%%%%%
%%%%%%%%%%%%%%%%%%%%%%%%%%%%%%%%%%%%%%%%%%%%%%%%%%%%%%%%%%%%%%%%%%%%%%%%%%%%%%%%%%%%%%%%%%%%%
\section{Related Work}\label{sec:relatedwork}
%%%%%%%%%%%%%%%%%%%%%%%%%%%%%%%%%%%%%%%%%%%%%%%%%%%%%%%%%%%%%%%%%%%%%%%%%%%%%%%%%%%%%%%%%%%%%
%%%%%%%%%%%%%%%%%%%%%%%%%%%%%%%%%%%%%%%%%%%%%%%%%%%%%%%%%%%%%%%%%%%%%%%%%%%%%%%%%%%%%%%%%%%%%

Since the first workshop in this field in 2004,
question answering over textual data with geospatial information has been studied by Geographic Information Retrieval researchers.
%(\eg see the GIR series of workshops).
%\footnote{\url{http://www.geo.uzh.ch/~rsp/gir/}}.
Relevant problems in this area include detecting place names (a special case of named entity recognition) and associated spatial natural language qualifiers in text and user queries, and disambiguating place names (a special case of named entity disambiguation). Two representative examples of systems where some of these issues have been studied are SPIRIT~\cite{DBLP:conf/sigir/JonesPRSSKW02} and STEWARD~\cite{DBLP:conf/gis/LiebermanSSS07}.

An important evaluation initiative for geographic information retrieval from multilingual text has been GeoCLEF.\footnote{\url{http://www.clef-initiative.eu/track/GeoCLEF}} From the 2008 version of GeoCLEF, the GiKiP pilot is complementary to our paper since it concentrated on answering geospatial questions in three languages (Portuguese, English and German) from Wikipedia~\cite{Santos2008GettingGA}.

Query processing for linked geospatial data has been an active field of research recently culminating in the definition of the OGC standard GeoSPARQL, an extension of SPARQL with a vocabulary, datatypes and functions for expressing geospatial queries over linked data.
There has also been substantial activity in the implementation of query processing systems such as Strabon~\cite{DBLP:conf/semweb/KyzirakosKK12} and Ontop-spatial~\cite{DBLP:conf/semweb/BeretaK16}, which both support GeoSPARQL.
The work of the present paper goes beyond these query processors to offering question answering services over linked geospatial data, i.e.\ supporting queries expressed in natural language.

The work by Younis et al.~\cite{DBLP:conf/giscience/YounisJTA12} is most closely related to our work since it presents a system for answering geospatial questions over DBpedia.
The system is based on a PostGIS\footnote{\url{http://postgis.net/}} database containing precise geospatial information of features in the United Kingdom provided by Ordnance Survey, a spatial index of DBpedia resources
built using their point coordinates, and a SPARQL endpoint storing the DBpedia dataset.
The three classes of questions considered are proximity
\qq{Find churches within 1~km of the River Thames}), crossing (\eg \qq{Find the mouths of the rivers that cross Oxford}) and containment (\eg \qq{Find churches in Manchester}). As we will see in Section ~\ref{sec:geopatial}, these kinds of questions are a subset of the ones that can be handled by GeoQA.
Younis et al.\cite{DBLP:conf/giscience/YounisJTA12} informally discusses the techniques that can be used to answer such questions, however, the system and dataset are not provided. %but makes no system available with which we could compare our GeoQA. In addition,\cite{DBLP:conf/giscience/YounisJTA12} makes no dataset available on which other approaches like ours can be evaluated. 
Finally, the discussion in the paper pays some attention to the quality of returned answers.

%but performance issues are not studied.

Grutter et al.~\cite{DBLP:journals/tgis/GrutterPW17} explore the use of DBpedia and Geonames for answering topological queries involving administrative divisions of Switzerland
and Scotland (since the authors are very familiar with the administrative geographies of these two countries). The paper contains a detailed discussion
of quality issues in linked geospatial data and especially the two linked data sources used by the authors (e.g., incompleteness, inconsistency of data etc.).
Finally, the paper considers queries for neighbouring and containing/contained administrative divisions, and measures precision and recall when only one of 
datasets or both linked datasets are used. 

Hamzaie et al.~\cite{DBLP:conf/agile/HamzeiLVB0T19} analyzes the natural language question-answer dataset MS MARCOV2.1 ~\cite{DBLP:conf/nips/NguyenRSGTMD16} which contains questions posed to the Bing search engine and human-generated answers. They concentrate on place-related questions of this dataset and define a set of patterns that can be used to characterize semantically questions and their answers. They also present a deeper understanding of the dataset using techniques based on word embeddings and clustering.

Tang and Mooney in~\cite{DBLP:conf/ecml/TangM01} presents an inductive logic programming approach for learning a semantic parser and applies its techniques to two areas, one of which is querying geospatial databases. They have experimented with a dataset consisting of 1000 Prolog facts from the U.S. Geography domain, and have also developed a corpus of 880 natural language questions and their corresponding logical queries in Prolog.\footnote{\url{http://www.cs.utexas.edu/users/ml/nldata/geoquery.html}} A part of this corpus is used to train the semantic parser developed by the authors.

As it is already mentioned, in the area of QA there is currently no engine that deals with geospatial questions like GeoQA. From the existing systems, PowerAqua needs to be mentioned in our context since it also assumes that questions will be answered from many datasets or ontologies~\cite{DBLP:journals/semweb/LopezFMS12}.

% The problem of temporal QA has also been studied in the literature, most recently by~\cite{DBLP:conf/cikm/JiaARSW18}.

% Finally, another related paper is~\cite{both2014get} where the notion of geospatial motives is presented as a way of aggregating geospatial information for search purposes.

\section{Constructing a Gold Standard Geospatial Data Set}\label{sec:GeoData}
%%%%%%%%%%%%%%%%%%%%%%%%%%%%%%%%%%%%%%%%%%%%%%%%%%%%%%%%%%%%%%%%%%%%%%%%%%%%%%%%%%%%%%%%%%%%%

In this section we discuss how to construct a gold standard geospatial dataset by interlinking DBpedia, OpenStreetMap and the GADM dataset of global administrative areas. Since DBpedia contains very limited geospatial information (e.g., latitude/longitude pairs, qualitative information via predicates such as {\tt dbo:Country}), we enrich DBpedia with quantitative geospatial information (i.e., geometries) by interlinking it with OSM and GADM.

GADM is a dataset containing information about administrative divisions of various countries and their boundaries. GADM 3.4 (released on May 2018) contains information about 386,735 administrative areas. Particularly, the multi-polygon for each administrative area is provided along with a set of qualitative data, including its  name and variant names. As the already existing linked data form of GADM~\cite{salas2011finding} was based on very old version of GADM (and its representation was not based on the GeoSPARQL standard ) we created a new linked data form of GADM from the available shapefiles using the tool GeoTriples\footnote{\url{http://geotriples.di.uoa.gr}}. We have the data from release version 2.8 (released on November 2015 ). For the purposes of this paper, we have only used GADM data from the United Kingdom (England, Scotland, Wales and Northern Ireland) and Ireland. The graphical representation\footnote{\url{http://geoqa.di.uoa.gr/images/gadm_ontology.png}} and the RDF/XML form\footnote{\url{http://geoqa.di.uoa.gr/assets/GADM.owl}} of the GADM ontology used are publicly available. Throughout the paper we use the prefix {\tt gadmr:} instead of {\tt \url{http://www.app-lab.eu/gadm}} for resources in the GADM data, and {\tt gadmo:} for {\tt \url{http://www.app-lab.eu/gadm/ontology}} for resources in the GADM ontology.

OSM is a collaborative project to create a free editable map of the world. It contains information about various features like rivers, lakes, cities, roads, points of interest (e.g., museums, restaurants and schools) etc. The geometries of these features can be points, lines or polygons. In addition to the geometry of a feature, OSM contains useful information such as name, feature class and layer. OSM data can be obtained in various formats. The first project to transform OSM data into RDF was LinkedGeoData~\cite{DBLP:conf/semweb/AuerLH09}. Currently, this project does not provide an up-to-date version of OSM data that we could use for our study. For this reason, we had to repeat some of the work presented in~\cite{DBLP:journals/semweb/StadlerLHA12} and, by doing this, go beyond \cite{DBLP:journals/semweb/StadlerLHA12} in the way that we will explain below. In the rest of the paper we use the prefix {\tt osmr:} instead of {\tt http://www.app-lab.eu/osm} for resources in the OSM data, and {\tt osmo:} instead of {\tt http://www.app-lab.eu/osm/ontology} for resources in the OSM ontology.

We obtained the OSM dataset in shapefile format from the company GEOFABRIK~\footnote{\url{http://download.geofabrik.de/europe.html}} and converted it into RDF using the tool GeoTriples. These shapefiles contain data available on date 30th August 2017. Like GADM, we have restricted our attention to the United Kingdom and Ireland. We designed a new ontology for OSM data which closely models the data in the shapefiles and made it publicly available in graphical format\footnote{\url{http://sites.pyravlos.di.uoa.gr/dragonOSM.svg}} and in RDF/XML format\footnote{\url{http://pyravlos-vm5.di.uoa.gr/osm.owl}}. The ontology uses the GeoSPARQL vocabulary to model the geometries of various OSM features. Note that OSM does \textit{not} have detailed administrative boundaries of various countries, hence we retrieve this information from GADM.

DBpedia is one of the most popular knowledge graphs derived from Wikipedia and its ontology which we use in the paper is publicly available\footnote{\url{http://mappings.dbpedia.org/server/ontology/classes/ }}. Throughout the paper we use the prefix dbo: instead of \code{http:dbpedia.org/ontology} for resources in the DBpedia ontology, and dbr: instead of \code{http://dbpedia.org/resource} for resources in the DBpedia knowledge graph.
Interlinking of GADM and OSM with DBpedia allows us to answer geospatial questions that cannot be answered by any of the datasets in isolation. For example, the question``Which of the English counties that border Greater Manchester has the highest percentage of ethnic Asians?'' can only be answered by consulting GADM to find the counties that border Greater Manchester, and then DBpedia to find the percentage of various ethnic groups in these counties. Also, the question ``Which Greek politicians are graduates of a university located in a Greek island belonging to the region of Western Greece?'' can be answered only by consulting all three datasets.

\subsection{Interlinking GADM with DBpedia}
The interlinking of GADM with DBpedia was done as follows. 
For each administrative area mentioned in GADM we obtain the DBpedia resource which has the same label, by using DBpedia SPARQL endpoint. The few remaining GADM resources were mapped manually.
%We get the name of an administrative area from GADM and we query the DBpedia SPARQL endpoint to get the DBpedia resource having same label with the resource of GADM. Then we link these two resources using \texttt{owl:sameAs}.
This procedure resulted in most of the GADM resources being linked. The remaining ones were linked manually. Table \ref{tab:gadm_interlink} shows the relevant numbers.

\begin{table*}
	\small
    \centering
    \caption{Interlinking GADM with DBpedia}
     \begin{tabular}{ | c | c| c | c | }
            \hline
            \bf{Country} & \bf{Total} & \bf{Linked} & \bf{Linked} \\
                         & \bf{entities} & \bf{automatically} & \bf{manually}\\
            \hline
            \bf{UK} & 197 & 164 & 33 \\
            \hline
            \bf{Ireland} & 27 & 17 & 10 \\
            \hline
    \end{tabular}
    \label{tab:gadm_interlink}
\end{table*}
\sloppypar{}
\subsection{Interlinking of OSM with DBpedia}
The task of interlinking OSM with DBpedia had some interesting challenges. First of all, we manually identified classes that have the same or very similar label in DBpedia and OSM. These classes are: {\tt Airport, Bank, Beach, Building, Canal, Castle, Cemetery, Church, City, College, Dam, Forest, Fort, Glacier, Golfcourse, Hospital, Hotel, Island, Library, Lighthouse, Locality, Memorial, Mine, Monument, Mosque, Museum, Park, Place, Prison, RailwayStation, Region, Restaurant, River, Road, School, Stadium, Stream, Temple, Theatre, Tower, Town, Tram, University, Village, Volcano, Watertower, Watermill, Windmill,} and {\tt Zoo}. Then, interlinking was performed on a class-by class basis using the tool Silk~\cite{DBLP:conf/www/VolzBGK09}. The OSM data is stored in a Strabon endpoint and the online DBpedia endpoint is used for the DBpedia data. The labels of the entities and the spatial distance of their geometries were considered equally for matching. In other words, we use the formula
$(S(x,y) + MinDist(x,y))/2 = 1.0$
where
\begin{itemize}
    \item $x$ and $y$ are the instances considered for matching in OSM and DBpedia respectively.
    \item $S(x,y)$  is the Levenshtein string similarity measure between the labels of $x$ and $y$. The threshold considered for string similarity is 85\%.
    %If string similarity is up to 85\% then S(x,y) = 1.0 in the above formula.
    \item $MinDist(x,y)$ is the minimum Euclidean distance between the geometries of $x$ and $y$. After experimenting with different number of threshold values for Euclidean distance, we finalized the threshold to 1 kilometer.
    %Minimum spatial distance between x and y must be up to threshold distance than only we can consider MinDist(x,y) = 1.0 in the above formula.
\end{itemize}

Table \ref{tab:osm_interlinked} gives the number of instances of the various classes in both datasets, as well as the number of instances that were interlinked. The DBpedia instances have been selected by retrieving only the resources that have coordinates falling inside the minimum bounding rectangles of the geometries of the United Kingdom and Ireland. As it is expected, some classes have many more instances in one of the datasets. For example, the class {\tt Restaurant} has 24055 instances in the subset of OSM that we consider and only 152 instances in DBpedia. Also, some of the classes having the same label are at different places in the class hierarchies of the two datasets. For example, the class {\tt Building} is the parent class of {\tt Restaurant, Hotel, Hospital, Museum} etc. in the DBpedia ontology, while it does not have any subclasses in the hierarchy of the OSM ontology, so we interlink instances of the subclasses of {\tt Building} with the instances of corresponding classes of OSM. Similarly, {\tt Road} has subclasses that we consider in OSM ontology,  while it does not have any subclasses in DBpedia. Naturally, when a class had zero instances in one or both datasets (e.g., {\tt Beach} in the DBpedia subset we consider and {\tt Glacier} in both datasets) then the class does not participate in the interlinking and does not appear in Table \ref{tab:osm_interlinked}. Finally, we would like to mention that we found many misclassified instances in DBpedia in contrast to the other two datasets; this has also been pointed out in~\cite{DBLP:journals/semweb/StadlerLHA12}.

Let us now comment on some rows of Table \ref{tab:osm_interlinked} where there is an unexpectedly big difference in the number of instances in OSM and DBpedia for the same class. Let us take for example the class {\tt Airport}. Unfortunately, the freely available OSM shapefiles for the United Kingdom and Ireland, provided by GEOFABRIK~\footnote{\url{http://download.geofabrik.de/europe.html}}, contain only 7 airports (not even Heathrow airport of London is included!). On the contrary, DBpedia has a rather large number resources classified as airports. In some cases, these are wrongly classified e.g., {\tt dbr:Brahan\_Castle}, a castle, is wrongly classified as {\tt dbo:Airport}.  It is also interesting to consider the row for class {\tt River}. There are many more instances of {\tt River} in OSM than in DBpedia because OSM has a different entry for each of the segments/polygons making up a river in its full length. The same issue exists for the classes {\tt Canal} and {\tt Stream}. This is the reason that the number of total interlinked instances is bigger than the cardinality of the intersection of the two datasets for classes like {\tt River}. Finally, another reason for the difference in instances between the same classes in DBpedia and OSM is the nature of the domain of interest in the datasets. For example, DBpedia has information about only 1339 hotels in its full dataset of which 212 hotels are in the United Kingdom and Ireland. The corresponding number of hotels in OSM is 9819 hotels as we see from the Table~\ref{tab:osm_interlinked}. In a similar way, the class {\tt Restaurant} in DBpedia has few instances compared to OSM.

After completing the interlinking with Silk, there were some entities that were not linked. These were checked and linked, if appropriate, manually.
For matches below 100 all matching pairs were checked manually for correctness. For larger numbers of matching pairs, we checked manually 100 random pairs and found them all to be correct. So, we conclude that our matching process is very accurate.
% and some of the class labels has zero entities linked e.g. college. To improve on this we manually checked data from both the sources. We found that a main reason all the entities could not be linked was the way labels are written in both data sources. E.g., “The Checkers (restaurant)” is label in DBpedia while “Chequers” is one of the label in OSM. Next step we have taken is we separated the entities from DBpedia that were not interlinked and using java code we have listed all the OSM instances of the same class labels that are within respected threshold distance of the class label from the DBpedia instance. From this shortlisted nearby instances, we compared the labels manually and interlinked if they are matching.

Comparing the interlinking of OSM and DBpedia that we have done with the interlinking done in LinkedGeoData~\cite{DBLP:journals/semweb/StadlerLHA12}, we can see that we have interlinked instances belonging to many more classes. The OSM dataset in the case of LinkedGeoData is stored using Virtuoso which has support only for point geometries. Therefore, no queries involving complex geometries can be done, and the interlinked resources in the case of LinkedGeoData are limited to OSM nodes.

\begin{table}[]
\caption{Interlinking OSM with DBpedia}
\begin{tabular}{| m{0.25\textwidth} | m{0.1\textwidth} | m{0.1\textwidth} | m{0.15\textwidth} | m{0.15\textwidth} | m{0.1\textwidth} |}
\hline
\textbf{Class} & \textbf{No. of Instances in OSM} & \textbf{No. of Instances in DBpedia} & \textbf{Interlinked Instances} & \textbf{Interlinked Instances (semi automatically)} & \textbf{Total Interlinked Instances} \\ \hline
Airport & 7 & 815 & 1 & 5 & 6  \\ \hline
Bank & 7621 & 29 & 1 & 2 & 3  \\ \hline
Canal & 7902 & 167 & 2171 & 920 & 3091 \\ \hline
Castle & 1357 & 486 & 161 & 36 & 197 \\ \hline
City & 86 & 101 & 45 & 18 & 63 \\ \hline
College & 1529 & 38 & 0 & 2 & 2 \\ \hline
Dam & 330 & 26 & 1 & 3 & 4 \\ \hline
Hospital & 2352 & 537 & 244 & 149 & 393 \\ \hline
Hotel & 9819 & 212 & 73 & 81 & 154 \\ \hline
Island & 2477 & 750 & 219 & 138 & 357 \\ \hline
Library & 3635 & 119 & 47 & 25 & 72 \\ \hline
Lighthouse & 423 & 39 & 9 & 14 & 23 \\ \hline
Monument & 2108 & 38 & 5 & 3 & 8 \\ \hline
Museum & 2313 & 933 & 327 & 219 & 546 \\ \hline
Park & 54830 & 382 & 252 & 103 & 355 \\ \hline
Prison & 207 & 199 & 28 & 119 & 137 \\ \hline
Railway Station & 3932 & 45 & 0 & 0 & 0 \\ \hline
Region & 13 & 151 & 0 & 0 & 0 \\ \hline
Restaurant & 24058 & 152 & 31 & 30 & 61 \\ \hline
River & 52897 & 785 & 4342 & 237 & 4579 \\ \hline
School & 33217 & 5556 & 2683 & 691 & 3374 \\ \hline
Stadium & 799 & 687 & 120 & 78 & 198 \\ \hline
Stream & 240293 & 470 & 885 & 265 & 1150 \\ \hline
Theatre & 1224 & 86 & 19 & 33 & 52 \\ \hline
Tower & 2373 & 35 & 0 & 0 & 0 \\ \hline
Town & 1960 & 1066 & 132 & 18 & 150 \\ \hline
University & 2466 & 1099 & 169 & 41 & 210 \\ \hline
Village & 15743 & 15346 & 4308 & 4087 & 8395 \\ \hline
\end{tabular}
\label{tab:osm_interlinked}
\end{table}
The GADM and OSM datasets as well as the interlinking dataset is publicly available on the Web site of the gold standard.\footnote{\url{http://geoqa.di.uoa.gr}} We will call this data part of the gold standard \textit{GeoData201}.
%%%%%%%%%%%%%%%%%%%%%%%%%%%%%%%%%%%%%%%%%%%%%%%%%%%%%%%%%%%%%%%%%%%%%%%%%%%%%%%%%%%%%%%%%%%%%
%%%%%%%%%%%%%%%%%%%%%%%%%%%%%%%%%%%%%%%%%%%%%%%%%%%%%%%%%%%%%%%%%%%%%%%%%%%%%%%%%%%%%%%%%%%%%
\section{Creating a gold standard set of geospatial questions}\label{sec:Benchmark}
%%%%%%%%%%%%%%%%%%%%%%%%%%%%%%%%%%%%%%%%%%%%%%%%%%%%%%%%%%%%%%%%%%%%%%%%%%%%%%%%%%%%%%%%%%%%%
%%%%%%%%%%%%%%%%%%%%%%%%%%%%%%%%%%%%%%%%%%%%%%%%%%%%%%%%%%%%%%%%%%%%%%%%%%%%%%%%%%%%%%%%%%%%%
To be able to evaluate the effectiveness of our query engine and compare it with other QA engines available, we have created a new benchmark set of 201 questions which we have collectively called \textit{GeoQuestions201}. The questions have been written by third-year students of the 2017-2018 Artificial Intelligence course in our department. The students were asked to target the above three data sources by imagining scenarios where geospatial information will be needed and could be provided by an intelligent assistant, and to propose questions with a
geospatial dimension that they considered \qq{simple} (a few examples of such questions were provided). The authors of the paper have then \qq{cleaned} the given set of questions and produced the SPARQL or GeoSPARQL queries that correspond to them assuming ontologies that describe the three data sources using the GeoSPARQL
vocabulary. The complete set of resources (data sources, ontologies, natural language questions and SPARQL/GeoSPARQL queries) is available on the Web at {\tt http://geoqa.di.uoa.gr}.

The questions in the benchmark GeoQuestions201 fall under the following
categories. For each one of the categories, we also comment on whether two major search engines (Google and Bing) can answer questions in this category.
\begin{enumerate}
\item \textit{Asking for the attribute of a feature}. For example, \qq{Where is Loch Goil located?} or \qq{What is length of River Thames?}. In GeoQA, these questions can be answered by posing a SPARQL query to DBpedia. Google and Bing both can also answer such questions precisely.

\item \textit{Asking whether a feature is in a geospatial relation with another feature}. For example, “Is Liverpool east of Ireland?”. The geospatial relation in this example question is a cardinal direction one (east of). Other geospatial relations in the set of questions include topological (borders) or distance (near or “at most 2km from"). In GeoQA, these questions are answered most of the time by using GADM and OpenStreetMap because the relevant qualitative geospatial knowledge is not present in DBpedia and/or the detailed geometries of features are needed for evaluating the geospatial relation of the question. Google and Bing both cannot answer such factoid questions; both can only return a list of relevant Web pages.

\item \textit{Asking for features of a given class that are in a geospatial relation with another feature}. For example, “Which counties border county Lincolnshire?” or “Which hotels in Belfast are at most 2km from George Best Belfast City Airport?”. The geospatial relation in the first example question is a topological one (border). As in the previous category, other geospatial relations in the set of questions include cardinal (e.g., southeast of) or distance (near or “at most 2km from” as in the second example question). In GeoQA, these questions can be answered by using not just DBpedia but also GADM and OpenStreetMap when the detailed geometries of features are needed for evaluating the geospatial relations. Google can also answer such questions precisely in many but not all cases (e.g., it can answer precisely the first and third questions but not the second). Bing cannot answer such questions precisely but gives list of relevant web pages.

Questions in this category might also have a second geospatial relation and a third feature which are used to further constrain the second feature. For example, ``Which restaurants are near Big Ben in London?'' or ``Which rivers cross London in Ontario?''. In the first question, we have also provided some more information about Big Ben although this might not have been necessary.\footnote{The authors of this paper are not aware of another Big Ben.}
In the second question, ``in Ontario'' is used to make clear that we are referring to the city London in Ontario, Canada not the more well-known city of London in England.\footnote{Boringly enough, London, Ontario is also crossed by a Thames river. We bet this is not how this river was called by native Indians in 1534 when Canada was discovered.}

\item \textit{Asking for features of a given class that are in a geospatial relation with any features of another class.} For example, ``Which churches are near castles?''. Arguably, this category of questions might not be useful unless one specifies a geographical area of interest; this is done by the next category of questions.

\item \textit{Asking for features of a given class that are in a geospatial relation with an unspecified feature of another class which, in turn, is in another geospatial relation with a feature specified explicitly.} An example of such a question is ``Which churches are near a castle in Scotland?''. Google and Bing both cannot answer such questions precisely. %Bing end up giving links to the pages that contains list of points of interest specified in the question.

\item \textit{The questions in this category are like the ones in Categories 3 to 5 above, but in addition, the thematic and/or geospatial characteristics of the features that are expected as answers (i.e., the features of the first class mentioned in the question) satisfy some further condition (e.g., numeric).} For example, \qq{Which mountains in Scotland have height more than 1000 meters?} or \qq{Which villages in Scotland have a population of less than 500 people?} or \qq{Is there a church in the county of Greater Manchester dedicated to St. Patrick?} or \qq{Which Greek restaurants in London are near Wembley stadium?}. In these examples, the extra attribute conditions may require GeoQA to consult all three data sources to find the answer to a question. Google can answer precisely the first, third and fourth example question, but not the second, since its
knowledge graph does not contain population information for villages in Scotland. Bing cannot answer any questions precisely but returns relevant links to points of interest.

\item \textit{Questions with quantities and aggregates.} For example, \qq{Which is the highest mountain in Ireland?} or \qq{Which hotel is the nearest to Old Trafford Stadium in Manchester?} or \qq{Which is the largest lake by area in Great Britain?} Questions with quantities but without aggregates have recently been studied by~\cite{DBLP:conf/semweb/HoIPBW19} in a non-geospatial setting. Interestingly, Google  and Bing both can answer all three example questions precisely. Note that questions in this class might also exhibit features of the previous two classes e.g., when a topological relation is involved or when the condition on an attribute refers to a quantity (e.g., height of a mountain). Such questions cannot be handled by QA engines as well as Google and Bing at the moment. For example, the question “Which is the largest county of England by population which borders Lincolnshire?” is answered incorrectly by Google (county Bristol is given as the answer) as well as by Bing.
\end{enumerate}
The list of benchmark questions is available publicly\footnote{\url{http://geoqa.di.uoa.gr/benchmarkquestions.html}}.
\section{Creating a Geospatial Question Answering Pipeline}\label{sec:geopatial}
%%%%%%%%%%%%%%%%%%%%%%%%%%%%%%%%%%%%%%%%%%%%%%%%%%%%%%%%%%%%%%%%%%%%%%%%%%%%%%%%%%%%%%%%%%%%%
%%%%%%%%%%%%%%%%%%%%%%%%%%%%%%%%%%%%%%%%%%%%%%%%%%%%%%%%%%%%%%%%%%%%%%%%%%%%%%%%%%%%%%%%%%%%%
We now present our approach to translating a natural language question into a GeoSPARQL query that can be executed on the union of the datasets presented in the previous section. For this, we build a geospatial question answering system using Qanary~\cite{DBLP:conf/icwe/DiefenbachSBC0A17} and Frankenstein~\cite{DBLP:conf/www/SinghRBSLUVKP0V18}.
%%%%%%%%%%%%%%%%%%%%%%%%%%%%%%%%%%%%%%%%%%%%%%%%%%%%%%%%%%%%%%%%%%%%%%%%%%%%%%%%%%%%%%%%%%%%%
%%%%%%%%%%%%%%%%%%%%%%%%%%%%%%%%%%%%%%%%%%%%%%%%%%%%%%%%%%%%%%%%%%%%%%%%%%%%%%%%%%%%%%%%%%%%%
\subsection{The Frankenstein Framework for Building QA Systems}\label{sec:qanary}
%%%%%%%%%%%%%%%%%%%%%%%%%%%%%%%%%%%%%%%%%%%%%%%%%%%%%%%%%%%%%%%%%%%%%%%%%%%%%%%%%%%%%%%%%%%%%
Qanary is a lightweight component-based QA methodology for the rapid engineering of QA pipelines~\cite{DBLP:conf/esws/BothDSSC016,DBLP:conf/icwe/BothSDL17}.
Frankenstein~\cite{DBLP:conf/www/SinghRBSLUVKP0V18} is the most recent implementation  of the ideas of Qanary; this makes it an excellent framework for developing reusable QA components and integrating them in QA pipelines. Frankenstein is built using the Qanary methodology developed by Both et al.~\cite{DBLP:conf/esws/BothDSSC016} and uses standard RDF technology to wrap and integrate existing standalone implementations of state-of-the-art components that can be useful in a QA system.
 The Qanary methodology is driven by the knowledge available for describing the input question and related concepts during the QA process. Frankenstein uses an extensible and flexible vocabulary \cite{DBLP:conf/esws/SinghBDSC016} for data exchange between the different QA components. This vocabulary establishes an abstraction layer for the communication of QA components.
While integrating components using Frankenstein, all the knowledge associated with a question and the QA process is stored in a process-independent knowledge base using the vocabulary. 
%Therefore, developers can reuse existing components for creating their own QA pipelines. 
Each component is implemented as an independent micro-service implementing the same RESTful interface. During the start-up phase of a QA pipeline, a service registry is automatically called by all components.
As all components are following the same service interface and are registered to a central mediator, they can be easily activated and combined by developers to create different QA systems.
%%%%%%%%%%%%%%%%%%%%%%%%%%%%%%%%%%%%%%%%%%%%%%%%%%%%%%%%%%%%%%%%%%%%%%%%%%%%%%%%%%%%%%%%%%%%%
%%%%%%%%%%%%%%%%%%%%%%%%%%%%%%%%%%%%%%%%%%%%%%%%%%%%%%%%%%%%%%%%%%%%%%%%%%%%%%%%%%%%%%%%%%%%%
\subsection{GeoQA: A Geospatial QA System}\label{sec:GeoQA}
\begin{figure}
\begin{center}
\includegraphics[height=7.6cm,width=1.0\textwidth,keepaspectratio]{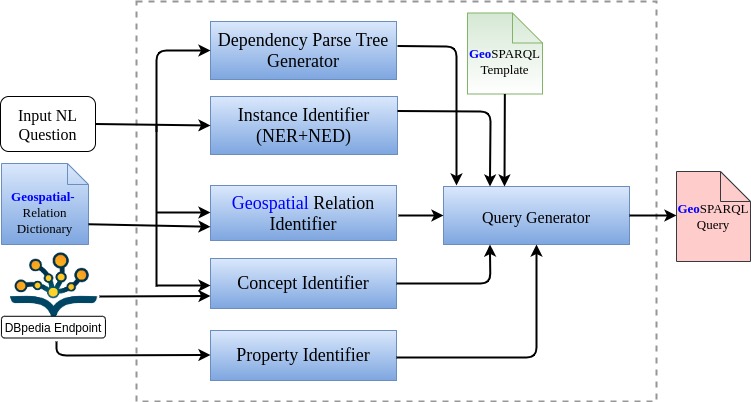}
\end{center}
\caption{The conceptual architecture of the GeoQA system}
\label{fig:geoqa_arch}
\end{figure}

\begin{figure}
\begin{center}
\includegraphics[height=8cm,width=2.0\textwidth,keepaspectratio]{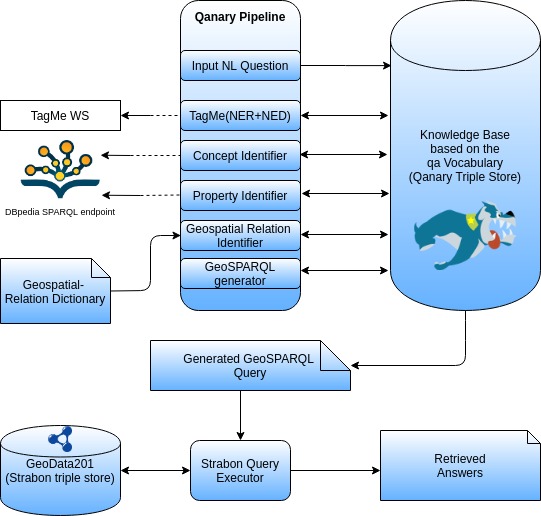}
\end{center}
\caption{Architecture of Implementation of GeoQA}
\label{fig:system_arch}
\end{figure}

%%%%%%%%%%%%%%%%%%%%%%%%%%%%%%%%%%%%%%%%%%%%%%%%%%%%%%%%%%%%%%%%%%%%%%%%%%%%%%%%%%%%%%%%%%%%%
%%%%%%%%%%%%%%%%%%%%%%%%%%%%%%%%%%%%%%%%%%%%%%%%%%%%%%%%%%%%%%%%%%%%%%%%%%%%%%%%%%%%%%%%%%%%%
In our work, we leverage the power of the Frankenstein framework to create QA components which collectively implement the geospatial QA pipeline of GeoQA.
The QA process of GeoQA uses the following modules implemented as components in the Frankenstein framework: dependency parse tree generator, concept identifier, instance identifier, geospatial relation identifier, property identifier, SPARQL/GeoSPARQL query generator and SPARQL/GeoSPARQL query executor.
Our components are fully integrated in the Frankenstein ecosystem and can be reused to implement geospatial features in other QA systems, as our implementation is not monolithic like the implementation of many other QA systems~\cite{DBLP:conf/esws/DubeyDSHL16,DBLP:conf/esws/LopezMU06,DBLP:conf/www/UngerBLNGC12}.
GeoQA takes as input a question in natural language (currently only English is supported) and the three linked geospatial datasets presented in Section~\ref{sec:GeoData}, and produces one or more answers that are resources of the given datasets.
Question answering is performed by translating the input question to a set of SPARQL or GeoSPARQL queries, ranking these queries, and executing the top ranked query over two endpoints using the SPARQL {\tt SERVICE} keyword. For DBpedia, we use its public Virtuoso endpoint\footnote{\url{http://dbpedia.org/sparql}} while for GADM, OSM and their interlinking dataset we use a Strabon endpoint.
%\footnote{In our current experimental prototype, the endpoint at \footnote{Link is not given due to anonymity requirement.} is being used.}
In Figure~\ref{fig:geoqa_arch}, we present the conceptual view of the implemented GeoQA system and Figure~\ref{fig:system_arch} presents how our system is actually implemented. The various components of GeoQA are discussed below.

%%%%%%%%%%%%%%%%%%%%%%%%%%%%%%%%%%%%%%%%%%%%%%%%%%%%%%%%%%%%%%%%%%%%%%%%%%%%%%%%%%%%%%%%%%%%%
\paragraph{Dependency parse tree generator}
%%%%%%%%%%%%%%%%%%%%%%%%%%%%%%%%%%%%%%%%%%%%%%%%%%%%%%%%%%%%%%%%%%%%%%%%%%%%%%%%%%%%%%%%%%%%%
This component carries out part-of-speech tagging and generates a dependency parse tree for the input question using the Stanford CoreNLP software. The dependency parse tree is produced in CoNLL-U format~\cite{DBLP:conf/lrec/NivreMGGHMMPPST16}.
%based on the Universal Dependencies parts-of-speech (POS) and dependency tag sets.
%%%%%%%%%%%%%%%%%%%%%%%%%%%%%%%%%%%%%%%%%%%%%%%%%%%%%%%%%%%%%%%%%%%%%%%%%%%%%%%%%%%%%%%%%%%%%
\paragraph{Concept identifier}
%%%%%%%%%%%%%%%%%%%%%%%%%%%%%%%%%%%%%%%%%%%%%%%%%%%%%%%%%%%%%%%%%%%%%%%%%%%%%%%%%%%%%%%%%%%%%
The concept identifier module identifies the \textit{types of features} specified by the user in the input question and maps them to the corresponding classes in the DBpedia, GADM and OSM ontologies. We use the equivalent ontology-oriented term \textit{concept} for a feature type in this paper. For example, if the input question is \qq{Which restaurants are near Big Ben in London?}, then the term \qq{restaurants} is identified as a feature type and mapped to the class {\tt osmo:Restaurant} in the OSM ontology and {\tt dbo:Restaurant} in the DBpedia ontology. The matching classes are found using string matching on the labels of the classes (the Java library function  {\tt java.util.regex.Pattern.matcher()}is  used)together with lemmatization from Stanford CoreNLP and synonyms from Wordnet. In its final stage, the concept identifier annotates the appropriate node of the dependency parse tree with its results.
%%%%%%%%%%%%%%%%%%%%%%%%%%%%%%%%%%%%%%%%%%%%%%%%%%%%%%%%%%%%%%%%%%%%%%%%%%%%%%%%%%%%%%%%%%%%%
\paragraph{Instance identifier}
%%%%%%%%%%%%%%%%%%%%%%%%%%%%%%%%%%%%%%%%%%%%%%%%%%%%%%%%%%%%%%%%%%%%%%%%%%%%%%%%%%%%%%%%%%%%%
The next useful information to be identified in an input question is the \textit{features} mentioned. These can be e.g., the country Ireland or the city Dublin or the river Shannon etc. We use the equivalent ontology-oriented term \emph{instance(s)} for features in this paper. Once instances are identified, they are mapped to DBpedia, OSM or GADM resources using the entity recognition and disambiguation tool TagMeDisambiguate ~\cite{DBLP:conf/cikm/FerraginaS10}\footnote{\url{https://tagme.d4science.org/tagme/}}.

 Let us explain why we selected TagMeDisambiguate. We have considered the tools that have been used in ~\cite{DBLP:conf/www/SinghRBSLUVKP0V18} for entity recognition and disambiguation. We have tested these tools over \textit{GeoQuestions201}. It is to keep in mind that these tools may or may not be targeted for short text and text documents. We have used web services of these tools to annotate the \textit{GeoQuestions201}. Table~\ref{tab:ned analysis} shows information on these tools with how they perform over \textit{GeoQuestion201}. Column accuracy in Table~\ref{tab:ned analysis} is percentage of questions from the 201 we have that are correctly annotated with entities and their instances from DBpedia, Wikipedia or Wikidata. 

 The annotations from each tool were verified manually. As can be seen in Table~\ref{tab:ned analysis}, TagMeDisambiguate annotates most of the questions correctly compared to other tools.

\begin{table}
\small
\begin{center}
\caption{Accuracy of various named entity recognizer and disambiguator over Geoquestion201}
\begin{tabular}{ | m{0.4\textwidth} | m{0.3\textwidth} | m{0.2\textwidth} | }
\hline
\textbf {NER+NED tool} & \textbf{Accuracy over GeoQuestions201($\%$)} & \textbf{Disambiguate to}\\ %\textbf{Type of Target Text }
\hline
 StanfordNER~\cite{DBLP:conf/acl/FinkelGM05} + AIDA~\cite{DBLP:journals/pvldb/YosefHBSW11} & 80.0 & Wikipedia \\
 \hline
 DBpedia Spotlight~\cite{DBLP:conf/i-semantics/MendesJGB11} & 79.22 & DBpedia \\
 \hline
 StanfordNER~\cite{DBLP:conf/acl/FinkelGM05} + AGDISTIS~\cite{DBLP:conf/semweb/UsbeckNRGCAB14} & 81.50 & DBpedia \\
 \hline
 \textbf{TagMe~\cite{DBLP:conf/cikm/FerraginaS10}} & \textbf{89.5} & Wikipedia\\
 \hline
 MeaningCloud\footnote{\url{https://www.meaningcloud.com/}} & 67.0 & Wikipedia \\
 \hline
 TextRazor\footnote{\url{https://www.textrazor.com/}} & 76.5 & Wikipedia \\
 \hline
 Babelfly~\cite{DBLP:journals/tacl/0001RN14} & 40.5 & DBpedia \\
  \hline
 Entity-fishing\footnote{\url{https://nerd.readthedocs.io/en/latest/train.html}} & 84.5 & Wikipedia\\
 \hline
\end{tabular}
\label{tab:ned analysis}
\end{center}
\end{table}

We also search for resources in the OSM and GADM dataset that have the same label as the entity identified by the TagMeDisambiguate component, and add them to the list of identified instances. For illustration, consider the input question \qq{Which airports are in London?}.
The term \qq{London}\ is the identified instance (feature) and it is disambiguated to the wikipedia link and we get DBpedia resource {\tt dbr:London} by {\tt owl:sameAs} link from DBpedia Virtuoso endpoint\footnote{\url{http://dbpedia.org/sparql}}, and to
{\tt osmr:england/places/id/107775} and {\tt gadmr:administrativeUnit\_GBR\_adm2\_56} by our code. In its final stage, the instance identifier annotates the appropriate node of the dependency parse tree with its results.
%%%%%%%%%%%%%%%%%%%%%%%%%%%%%%%%%%%%%%%%%%%%%%%%%%%%%%%%%%%%%%%%%%%%%%%%%%%%%%%%%%%%%%%%%%%%%
\paragraph{Geospatial relation identifier}
%%%%%%%%%%%%%%%%%%%%%%%%%%%%%%%%%%%%%%%%%%%%%%%%%%%%%%%%%%%%%%%%%%%%%%%%%%%%%%%%%%%%%%%%%%%%%
Geospatial questions such as the ones targeted by GeoQA almost always include a qualitative geospatial relation such as ``borders'' or a quantitative one such as ``at most 2km from''. The current implementation supports the 14 geospatial relations shown on Table~\ref{tab:spatial relations}. 
These include some topological, some distance and some cardinal direction relations~\cite{DBLP:journals/gis/EgenhoferF91,DBLP:journals/vlc/Frank92,DBLP:conf/ssd/SkiadopoulosK01}.
Table~\ref{tab:spatial_relation_dictionary} gives a dictionary of the various synonyms for these relations that can appear instead of them in a question. The semantics of topological relations are as in the dimensionally extended 9-intersection model~\cite{DBLP:journals/isci/ClementiniF96}.
Qualitative spatial relations of proximity like ``close to", ``near" etc. are translated into (rather arbitrary) quantitative distance relations based on the concept identified earlier by the concept identifier (e.g.,  when asking for ``hotels near a place",  ``near'' is taken to mean at most 1 kilometer).
The semantics of cardinal direction relations are the usual ones i.e., a relation $A$ north of $B$ is given meaning by considering the bounding box of the reference region $B$ and the partition of the plane in nine areas that is induced by it~\cite{DBLP:conf/ssd/SkiadopoulosK01}. The same semantics are implemented by the Strabon system and its query language stSPARQL which is used as our back end geospatial RDF store~\cite{DBLP:conf/semweb/KyzirakosKK12}. GeoSPARQL does not support any cardinal direction functions or relations. Finally, Kreveld and Reinbacher~\cite{DBLP:journals/ijcga/KreveldR04} provide a more intuitive semantics of cardinal directions for polygons, but an implementation of this semantics is more expensive computationally than the semantics used in Strabon~\cite{DBLP:conf/semweb/KyzirakosKK12}.

Like the previous modules, this module first identifies geospatial relations in the input question, and then maps them to a spatial function of the GeoSPARQL or stSPARQL vocabulary, or a data property with a spatial semantics in the DBpedia ontology. As we have already discussed in the introduction, DBpedia contains limited explicit or implicit geospatial knowledge using latitude/longitude pairs, and properties such as {\tt dbp:northeast} for cardinal direction relations or class-specific properties such as {\tt dbo:city} (e.g., for class {\tt dbr:River}). GeoQA does not make use of quantitative geospatial information (i.e., latitude/longitude pairs) from DBpedia since we have more detailed geospatial knowledge in the form of polygons in the datasets GADM and OSM. However, it does makes use of qualitative geospatial knowledge from DBpedia expressed using the data properties just mentioned (although this knowledge is rather scarce as discussed in~\cite{DBLP:conf/esws/RegaliaJG16}).
As an example, for the question \qq{Which counties border Lincolnshire?}, the geospatial relation \qq{borders} is identified from the verbs in the dependency tree, and it is mapped to the spatial function \code{geof:sfTouches} of the GeoSPARQL vocabulary.

In its final stage, the geospatial relation identifier annotates the appropriate node of the dependency parse tree with its results. 
In the near future, GeoQA will cover all the prototypical spatial relations shown experimentally to correspond to natural language utterances by Egenhofer, Mark and their colleagues in~\cite{DBLP:journals/gis/ShariffEM98,mark1995topology,DBLP:conf/giscience/DubeE12}.

\begin{table}
\small
\begin{center}
\caption{Geospatial relation categories and relations}
\begin{tabular}{ | m{0.35\textwidth} | m{0.55\textwidth} |  }
\hline
\textbf {Category} & \textbf{Geospatial relation} \\
\hline
 Topological relations & \qq{within}, \qq{crosses}, \qq{borders} \\
\hline
 Distance relations & \qq{near}, \qq{at most $x$ units}, \qq{at least $x$ units} \\
 \hline
 Cardinal direction relations & \qq{north of}, \qq{south of}, \qq{east of}, \qq{west of}, \qq{northwest of}, \qq{northeast of}, \qq{southwest of}, and \qq{southeast of} \\
 \hline
\end{tabular}
\label{tab:spatial relations}
\end{center}
\end{table}
\begin{table}
\small
\begin{center}
\caption{Geospatial relations and their synonyms}
\begin{tabular}{ | m{0.3\textwidth} | m{0.6\textwidth} | }
\hline
\textbf {Geospatial relation} & \textbf{Synonyms in dictionary} \\
\hline
 within & in, inside, is located in, is included in \\
\hline
crosses & cross, intersect \\
\hline
near &  nearby, close to, around \\
\hline
borders & is/are at the border of, is/are at the outskirts of, at the boundary of\\
 \hline
 north of &  above of\\
 \hline
 south of &  below\\
 \hline
 east of &  to the right\\
 \hline
 west of &  to the left \\
 \hline
%=======================================================================================
\end{tabular}
\label{tab:spatial_relation_dictionary}
\end{center}
\end{table}
\begin{figure}[t]
\small
\begin{lstlisting} [frame=single,flexiblecolumns=t,caption={SPARQL/GeoSPARQL Query for Motivating Example 1},label={lst:limerick_example}]
Question: Which rivers cross Limerick?
SPARQL: 
select ?x
where {
		    ?x rdf:type dbo:River.
		    ?x dbo:city dbr:Limerick. 
}

(*\bfseries GeoSPARQL*):
PREFIX geo:  <http://www.opengis.net/ont/geosparql#>
PREFIX geof: <http://www.opengis.net/def/function/geosparql/>
PREFIX osmo: <http://www.app-lab.eu/osm/ontology#>

select ?x 
where { 
    		?x rdf:type osmo:River; 
    		   geo:hasGeometry ?xGeom. 
    		?xGeom geo:asWKT ?xWKT. 
    		
             gadmr:Limerick geo:hasGeometry ?iGeom.
    		?iGeom geo:asWKT ?iWKT.
    
    		FILTER(geof:sfCrosses(?xWKT, ?iWKT))
}
\end{lstlisting}
\end{figure}
%=======================================================================================
\begin{figure}[t]
\small
\begin{lstlisting} [frame=single,flexiblecolumns=t,caption={SPARQL/GeoSPARQL Query for Motivating Example 2},label={lst:count_example}]
(*\bfseries Question*): How many hospitals are there in Oxford?
(*\bfseries SPARQL*): 
select (count(distinct ?x) as ?total)
where {
    		?x rdf:type dbo:Hospital.
    		?x dbp:locatedIn dbr:Oxford. 
}

(*\bfseries GeoSPARQL*):
PREFIX geo:  <http://www.opengis.net/ont/geosparql#>
PREFIX geof: <http://www.opengis.net/def/function/geosparql/>
PREFIX osmo: <http://www.app-lab.eu/osm/ontology#>

select (count(distinct ?x)as ?total) 
where { 
            ?x rdf:type osmo:Hospital;
                geo:hasGeometry ?xGeom. 
            ?xGeom geo:asWKT ?xWKT. 

	         gadmr:Oxford geo:hasGeometry ?iGeom.
            ?iGeom geo:asWKT ?iWKT.

            FILTER(geof:sfWithin(?xWKT, ?iWKT))
}
\end{lstlisting}
\end{figure}
%=======================================================================================
%\begin{figure}[t]
%\small
%\begin{lstlisting} [frame=single,flexiblecolumns=t,caption={SPARQL/GeoSPARQL Query for Motivating Example 3},label={lst:location_example}]
%SPARQL: 
%select ?location
%where {
%		dbr:Big_Ben dbp:location ?location. 
%}
%\end{lstlisting}
%\end{figure}
%=======================================================================================
\begin{figure}
\small
\begin{lstlisting} [frame=single,flexiblecolumns=t,caption={GeoSPARQL Query for Motivating Example 3},label={lst:owlSameAs}]
(*\bfseries Question*):Which forest is near Manchester?
(*\bfseries GeoSPARQL*):
PREFIX owl:  <http://www.w3.org/2002/07/owl#>
PREFIX geo:  <http://www.opengis.net/ont/geosparql#>
PREFIX geof: <http://www.opengis.net/def/function/geosparql/>
PREFIX osmo: <http://www.app-lab.eu/osm/ontology#>
select ?x 
where { 
            ?x rdf:type osmo:Forest; 
                geo:hasGeometry ?xGeom. 
            ?xGeom geo:asWKT ?xWKT. 
            ?instance owl:sameAs dbr:Manchester; 
                geo:hasGeometry ?iGeom. 
            ?iGeom geo:asWKT ?iWKT. 
            FILTER(geof:distance(?xWKT,?iWKT,uom:metre) <= 5000) 
}
\end{lstlisting}
\end{figure}
%=======================================================================================
\begin{figure}[t]
\small
\begin{lstlisting} [frame=single,flexiblecolumns=t,caption={SPARQL/GeoSPARQL Query for Motivating Example 4},label={lst:catsix}]
(*\bfseries Question*):What is the length of the river that crosses Limerick?
(*\bfseries SPARQL*):
select ?length 
where { 
            ?x rdf:type    dbo:River.
            ?x dbo:city    dbr:Limeric.
            ?X dbp:length  ?length.
}
(*\bfseries GeoSPARQL*):
PREFIX geo:  <http://www.opengis.net/ont/geosparql#>
PREFIX geof: <http://www.opengis.net/def/function/geosparql/>
PREFIX osmo: <http://www.app-lab.eu/osm/ontology#>

select ?length 
where { 
       SERVICE <http://pyravlos1.di.uoa.gr:8080/geoqa/Query>
       {
		     ?x rdf:type osmo:River; 
		     geo:hasGeometry ?xGeom;
		     owl:sameAs ?dbpediaLink.
		     ?xGeom geo:asWKT ?xWKT. 

		     gadmr:Limerick geo:hasGeometry ?iGeom.
		     ?iGeom geo:asWKT ?iWKT.
            
	    	 FILTER(geof:sfCrosses(?xWKT, ?iWKT))
	    }
		SERVICE <http://dbpedia.org/sparql> 
		{
		    ?dbpediaLink dbp:length ?length
		}
}
\end{lstlisting}
\end{figure}
%%%%%%%%%%%%%%%%%%%%%%%%%%%%%%%%%%%%%%%%%%%%%%%%%%%%%%%%%%%%%%%%%%%%%%%%%%%%%%%%%%%%%%%%%%%%%
\paragraph{Property Identifier}
%%%%%%%%%%%%%%%%%%%%%%%%%%%%%%%%%%%%%%%%%%%%%%%%%%%%%%%%%%%%%%%%%%%%%%%%%%%%%%%%%%%%%%%%%%%%%
The property identifier module identifies \textit{ attributes of types of features} and \textit{attributes of features} specified by the user in input questions and maps them to corresponding properties in DBpedia. To answer questions like \qq{Which rivers in Scotland have more than 100 km length?} or \qq{Which mountains in Scotland have height more than 1000 meters?}, we need information about length of rivers and height of the mountains in addition to their geometry from OSM. This information is not present in OSM but we can retrieve this information from DBpedia. We use Table ~\ref{tab:dbpedia_properties} and ~\ref{tab:dbpedia_value_properties} for this task. The identified concept from the concept identifier module is used to search Table~\ref{tab:dbpedia_properties} to get \code{dbp:height} and \code{dbp:length} in the case of example questions mentioned before. In case of question \qq{Which Greek restaurants in London are near Wembley stadium?}, it is to be inferred that Greek should be a cuisine in context of restaurants and we need to check all the possible values of properties for the identified concept. We achieve this with the use of Table ~\ref{tab:dbpedia_value_properties}. We stress that Tables ~\ref{tab:dbpedia_properties} and ~\ref{tab:dbpedia_value_properties} contain only examples of classes, properties and values that are of interest to the example questions. In reality the tables contain 11,392 and 2,61,455 entries respectively and cover all the DBpedia classes of Table ~\ref{tab:osm_interlinked}. These tables have been generated querying DBpedia and stored in different files with their class names. In similar manner for question like \qq{What is the total area of Northern Ireland?} we query DBpedia endpoint to retrieve property \code{dbp:areaKm} that is present in DBpedia for identified instance \code{dbr:Northern\_Ireland}. We use string similarity measures while searching Table~\ref{tab:dbpedia_properties} and pattern matching while searching Table~\ref{tab:dbpedia_value_properties}. In its final stage, the property identifier annotates the appropriate node of the dependency parse tree with its results.

\begin{table}
\small
\begin{center}
\caption{DBpedia Properties }
\begin{tabular}{ | m{0.13\textwidth} | m{0.52\textwidth} | m{0.22\textwidth} | }
\hline
\textbf{DBpedia Class} & \textbf{DBpedia Property} & \textbf {Label of property} \\
\hline
Mountain & http://dbpedia.org/property/height & Height \\
\hline
Mountain & http://dbpedia.org/property/elevation & elevation \\
\hline
Mountain & http://dbpedia.org/property/parentPeak & Parent peak \\
\hline
River &	http://dbpedia.org/property/length & length \\
\hline
River &	http://dbpedia.org/property/name & name \\
\hline
River &	http://dbpedia.org/property/dischargeLocation &	discharge location \\
\hline
River &	http://dbpedia.org/property/mouth &	Mouth \\
 \hline
\end{tabular}
\label{tab:dbpedia_properties}
\end{center}
\end{table}

\begin{table}
\small
\begin{center}
\caption{DBpedia Properties and Values}
\begin{tabular}{ | m{0.2\textwidth} | m{0.47\textwidth} | m{0.2\textwidth} | }
\hline
\textbf{DBpedia Class} & \textbf{DBpedia Property} & \textbf {Value of property} \\
\hline
Restaurant & http://dbpedia.org/ontology/cuisine	 & Asian Cuisine \\
\hline
Restaurant & http://dbpedia.org/ontology/cuisine	& Italian,pizzeria \\
\hline
Restaurant & http://dbpedia.org/ontology/cuisine	& Italian, Greek, French, Spanish, and Creole table delicacies\\
\hline
\end{tabular}
\label{tab:dbpedia_value_properties}
\end{center}
\end{table}
%%%%%%%%%%%%%%%%%%%%%%%%%%%%%%%%%%%%%%%%%%%%%%%%%%%%%%%%%%%%%%%%%%%%%%%%%%%%%%%%%%%%%%%%%%%%%
\paragraph{Query generator}
%%%%%%%%%%%%%%%%%%%%%%%%%%%%%%%%%%%%%%%%%%%%%%%%%%%%%%%%%%%%%%%%%%%%%%%%%%%%%%%%%%%%%%%%%%%%%
This module creates a SPARQL or a GeoSPARQL query using handcrafted query templates. From gathering questions from  Google Trends and also studying the questions in our gold standard, we have identified the question patterns shown in Tables~\ref{Patterns} and ~\ref{Patternstwo}. In these tables C stands for ``concept'', I for ``instance'' , R for ``geospatial relation'' , P for ``property'' and N for ``Count of'' following the terminology we have introduced above. For each pattern, the tables give an example question and the corresponding GeoSPARQL and/or SPARQL query template. The query templates contain slots (strings starting with an underscore) that can only be identified when an example question is encountered and will be completed by the query generator (see below). 
\begin{table}
\small
\caption{Supported question patterns with examples and corresponding SPARQL/GeoSPARQL query templates for categories 1-5}
\begin{tabular}{ | m{0.1\textwidth} | m{0.2\textwidth} | m{0.8\textwidth} | }
\hline
\textbf {Pattern} & \textbf{Example natural language question} & {\textbf{Templates}} \\
\hline
 IP & Where is Emirates Stadium located? & \textbf{SPARQL}: \newline \code{select ?x
where \{
\newline	\_Instance \_Property ?x. 
\newline\}}\\
\hline
 CRI & Which rivers cross Limerick? & \textbf{SPARQL}: \newline select ?x
where \{
\newline		?x rdf:type \_Concept.
\newline		?x \_Relation \_Instance. 
\newline\}
\newline \textbf{GeoSPARQL} v1:\newline \code{select ?x 
where   \{ \newline ?x rdf:type \_Concept;
		   geo:hasGeometry ?xGeom. \newline
		    ?xGeom geo:asWKT ?xWKT. \newline
		    \_Instance geo:hasGeometry ?iGeom.
		    ?iGeom geo:asWKT ?iWKT.\newline
		    FILTER(\_Relation(?xWKT, ?iWKT))
\newline\}}
\newline \textbf{GeoSPARQL} v2: \newline \code{select ?x 
where \{ \newline ?x rdf:type \_Concept; \newline
		   geo:hasGeometry ?xGeom. \newline
		?xGeom geo:asWKT ?xWKT. \newline
		?instance owl:sameAs \_Instance; geo:hasGeometry ?iGeom.\newline
		?iGeom geo:asWKT ?iWKT.\newline
		FILTER(\_Relation(?xWKT, ?iWKT))
\newline\}} \\
\hline
 CRIRI & Which churches are close to the Shannon in Limerick? &  \code{select ?x
where \{ \newline
	    ?x rdf:type \_Concept; 
		   geo:hasGeometry ?xGeom. \newline
		?xGeom geo:asWKT ?xWKT. \newline
		\_Instance1 geo:hasGeometry ?i1Geom.\newline
		?i1Geom geo:asWKT ?i1WKT.\newline
		\_Instance2 geo:hasGeometry ?i2Geom.\newline
		?i2Geom geo:asWKT ?i2WKT. \newline
		FILTER(\_Relation1(?xWKT, ?i1WKT) \&\&  \_Relation2(?i1WKT, ?i2WKT) )
\newline\}}  
\\
 \hline
 CRC & Which restaurants are near hotels? &  \code{select ?x 
where \{ \newline
		?x rdf:type \_Concept1; 
		   geo:hasGeometry ?xGeom. \newline
		?xGeom geo:asWKT ?xWKT.\newline
		?y rdf:type \_Concept2; 
		   geo:hasGeometry ?yGeom. \newline
		?yGeom geo:asWKT ?yWKT.\newline
		FILTER(\_Relation(?xWKT, ?yWKT))
\newline\}}  
\\
 \hline
 CRCRI & Which restaurants are near hotels in Limerick? & \code{select ?x 
where \{ \newline
		?x rdf:type \_Concept1; 
		   geo:hasGeometry ?xGeom. \newline
		?xGeom geo:asWKT ?xWKT. \newline
		?y rdf:type \_Concept2;
		   geo:hasGeometry ?yGeom. \newline
		?yGeom geo:asWKT ?yWKT.\newline
		\_Instance geo:hasGeometry ?zGeom.\newline
		?zGeom geo:asWKT ?zWKT.\newline
		FILTER(\_Relation1(?xWKT, ?yWKT) \&\& \_Relation2(?xWKT, ?zWKT) \&\&\newline \_Relation2(?yWKT, ?zWKT)
\newline\}} 
\\
 \hline
 IRI & Is Hampshire north of Berkshire? &  \code{ASK
where \{ \newline
	    \_Instance1 geo:hasGeometry ?iGeom1. \newline
		?iGeom1 geo:asWKT ?iWKT1.\newline
		\_Instance2 geo:hasGeometry ?iGeom2.\newline
		?iGeom2 geo:asWKT ?iWKT2. \newline
		FILTER(\_Relation(?iWKT1, ?iWKT2))
\newline\}} 
\\
\hline
\end{tabular}
\label{Patterns}
\end{table}
%=======================================================================================
\begin{table}
\small
\caption{Supported question patterns with examples and corresponding SPARQL/GeoSPARQL query templates for categories 6,7}
\begin{tabular}{ | m{0.1\textwidth} | m{0.2\textwidth} | m{0.8\textwidth} | }
\hline
\textbf {Pattern} & \textbf{Example natural language question} & {\textbf{Templates}} \\
 \hline
 NCRI & How many hospitals are there in Oxford? & \textbf{SPARQL}: \newline \code{select (count(distinct ?x) as ?total)
where \{
\newline	?x rdf:type \_Concept.
\newline	?x \_Relation \_Instance. 
\newline\}}
\newline \textbf{GeoSPARQL} v1:\newline \code{select (count(distinct ?x) as ?total) 
where   \{ \newline ?x rdf:type \_Concept;
		   geo:hasGeometry ?xGeom. \newline
		?xGeom geo:asWKT ?xWKT. \newline
		\_Instance geo:hasGeometry ?iGeom.
		?iGeom geo:asWKT ?iWKT.\newline
		FILTER(\_Relation(?xWKT, ?iWKT))
\newline\}}
\newline \textbf{GeoSPARQL} v2: \newline \code{select (count(distinct ?x) as ?total) 
where \{ \newline ?x rdf:type \_Concept; \newline
		   geo:hasGeometry ?xGeom. \newline
		?xGeom geo:asWKT ?xWKT. \newline
		?instance owl:sameAs \_Instance; geo:hasGeometry ?iGeom.\newline
		?iGeom geo:asWKT ?iWKT.\newline
		FILTER(\_Relation(?xWKT, ?iWKT))
\newline\}} \\
 \hline
 PCRI & What is the length of the river that crosses Limerick? & \textbf{SPARQL}: \newline \code{select ?property
where \{
\newline		?x rdf:type \_Concept.
\newline		?x \_Relation \_Instance. 
\newline		?x \_Property ?property.
\newline\}} 
\newline \textbf{GeoSPARQL} :\newline \code{select ?property 
where   \{ \newline SERVICE <http://pyravlos1.di.uoa.gr:8080/geoqa/> \{ \newline ?x rdf:type \_Concept; 
		   geo:hasGeometry ?xGeom. \newline
		?xGeom geo:asWKT ?xWKT. \newline
		\_Instance geo:hasGeometry ?iGeom.\newline
		?iGeom geo:asWKT ?iWKT.\newline
		?x owl:sameAs ?dbpediaLink.
		FILTER(\_Relation(?xWKT, ?iWKT)) \} \newline
		SERVICE <http://dbpedia.org/sparql> \{ \newline
		?dbpediaLink \_Property ?property
		\}
\newline\}}\\
\hline
 PCRIRI & What is the name of the river that flows under the Queensway Bridge in Liverpool? & \textbf{GeoSPARQL} :\newline \code{select ?property 
where   \{ \newline SERVICE <http://pyravlos1.di.uoa.gr:8080/geoqa/> \{ \newline ?x rdf:type \_Concept; 
		   geo:hasGeometry ?xGeom. \newline
		?xGeom geo:asWKT ?xWKT. \newline
		\_Instance1 geo:hasGeometry ?i1Geom.\newline
		?i1Geom geo:asWKT ?i1WKT.\newline
		\_Instance2 geo:hasGeometry ?i2Geom.\newline
		?i2Geom geo:asWKT ?i2WKT. \newline
		?x owl:sameAs ?dbpediaLink. \newline
		FILTER(\_Relation1(?xWKT, ?i1WKT)) \&\&  \_Relation2(?i1WKT, ?i2WKT) \} \newline
		SERVICE <http://dbpedia.org/sparql> \{ \newline
		?dbpediaLink \_Property ?property
		\}
\newline\}}\\
 \hline
\end{tabular}
\label{Patternstwo}
\end{table}
\sloppypar{}
For each input question, the slots in the template are replaced by the query generator with the output of the previous modules, to generate a SPARQL or a GeoSPARQL query. For example, for the question \qq{What is length of the river that crosses Limerick?}, the identified pattern is PCRI. The question pattern is identified by searching the dependency parse tree in which the nodes have been annotated with the results of the concept, instance, property and geospatial relation identifier modules presented above. We walk through the parse tree with inorder traversal and identify the question pattern. If the question does not follow any of the patterns, a message is passed to the next component that no query has been generated.
The appropriate templates are selected from Table~\ref{Patterns} and ~\ref{Patternstwo}, their slots are filled with the resources identified earlier and the corresponding GeoSPARQL or SPARQL queries are generated. Here the concepts are \code{dbr:River} from DBpedia and \code{osmo:River} from OSM, the property is \code{dbp:length} from DBpedia, the instances are \code{gadmr:Limerick} from GADM and \code{osmo:irelandandnorthernireland/places/id/2518952} from OSM, and the geospatial relations are \code{dbo:city} from the DBpedia ontology and the GeoSPARQL function \code{geof:sfCrosses}. 

The row of Table~\ref{Patterns} for pattern CRI contains two GeoSPARQL queries (v1 and v2). The second query is for the case when the identified instance is a DBpedia resource for which geometry information is available in GADM or OSM. In addition to that, the rows for the patterns PCRI and PCRIRI in Table~\ref{Patternstwo} contains a service tag for a GeoSPARQL query in order to fetch information from two different endpoints to execute the query. This query is for the case when the identified instance is a DBpedia resource for which geometry information is available in OSM while attributes like \code{dbp:length} or \code{dbp:height} are in DBpedia. This is where the {\tt owl:sameAs} sentences produced by our interlinking process discussed in Section~\ref{sec:GeoData} are used. Listing ~\ref{lst:owlSameAs} and Listing ~\ref{lst:catsix} show examples of these cases. Similar templates exist for all the other patterns.

Because we want to increase recall, our strategy is to use more than one component of GeoData201 for answering a question. For example, for the question ``Which towns in England are east of Manchester?'' DBpedia gives us 3 answers (Glossop, Stallybridge and Hyde) while OSM gives us 1626 towns.
Our strategy for increasing precision is to have the query generator take into account class and property information from the ontologies of the three datasets. This is illustrated by the SPARQL query in Listing~\ref{lst:limerick_example} where we make use of the fact that the property {\tt dbo:city} is used in DBpedia to refer to the cities crossed by a river. To implement this strategy we keep a table with three columns which contains triples of the form domain-property-range for each property  in the dataset GeoData201. Some example rows can be seen in Table~\ref{tab:schema_table}. This approach has also been taken in~\cite{DBLP:conf/giscience/YounisJTA12}.

\begin{table}
\small
\begin{center}
\caption{domain-property-range Table}
\begin{tabular}{ | c | c | c | }
\hline
\textbf {Domain} & \textbf{Property} & \textbf{Range} \\
\hline
Airport & Within & http://dbpedia.org/ontology/city \\
\hline
Airport & Within & http://dbpedia.org/ontology/country \\
\hline
Airport & Within & http://dbpedia.org/ontology/county \\
\hline
River & Within & http://dbpedia.org/ontology/country \\
\hline
River & Within & http://dbpedia.org/ontology/county \\
\hline
River & Crosses & http://dbpedia.org/ontology/Bridge \\
\hline
River & Crosses & http://dbpedia.org/ontology/city \\
\hline
\end{tabular}
\label{tab:schema_table}
\end{center}
\end{table}

The last job of the query generator is to rank the generated queries. Query ranking is a crucial component of a question answering system. In the current version of GeoQA, we use a very simple heuristic for the ranking of generated queries based on the estimated selectivity of the generated queries. We compute the selectivity of a SPARQL or GeoSPARQL query taking into account only the triple patterns present in the query and using the formulas of~\cite{DBLP:conf/www/StockerSBKR08}. The generated query with the lowest selectivity is selected to be executed; in this way, we expect to generate more results to the user question.

\paragraph{Expressive Power of Patterns.} It is interesting to consider the expressive power of patterns in Tables~\ref{Patterns} and ~\ref{Patternstwo} by giving a corresponding binary first-order logic formula.\footnote{In the following formulas, we assume that identifiers (i.e., geographic features) are denoted by constants, concepts (i.e., classes of features) by unary predicates and geospatial relations by binary predicates.
Constants and predicates are denoted by capital letters while variables are denoted by lowercase letters. Variables are assumed to range over identifiers with the exception of variable $v$ in the case pf PCRI which ranges over values. The ``:'' symbol should be read as ``such that''.}
Questions following pattern IP can be written as $x: P(I,x)$. Questions following the CRI pattern can be written formally as $x:\ C(x) \wedge (\exists i) R(x,i)$. Questions following the pattern CRIRI can be written as
$x:\ C(x) \wedge (\exists i_1)(\exists i_2)(R_1(x,i_1) \wedge R_2(i_1,i_2))$. Questions following the pattern CRC can be written as $x:\ C_1(x) \wedge (\exists i)(C_2(i) \wedge R(x,i))$. Questions written as CRCRI can be written as $x:\ C_1(x) \wedge (\exists i_1)(\exists i_2)(R(x,i_1) \wedge C_2(i_2) \wedge R_2(i_1,i_2))$. Questions following pattern IRI can be written as $: R(I_1,I_2)$. Questions following the PCRI pattern can be written formally as $v: (\exists x)(C(x) \wedge R(x,I) \wedge P(I,v))$.

%%%%%%%%%%%%%%%%%%%%%%%%%%%%%%%%%%%%%%%%%%%%%%%%%%%%%%%%%%%%%%%%%%%%%%%%%%%%%%%%%%%%%%%%%%%%%
\paragraph{Query executor.}
%%%%%%%%%%%%%%%%%%%%%%%%%%%%%%%%%%%%%%%%%%%%%%%%%%%%%%%%%%%%%%%%%%%%%%%%%%%%%%%%%%%%%%%%%%%%%
The last module executes the top-ranked SPARQL or GeoSPARQL query against a Strabon~\cite{DBLP:conf/semweb/KyzirakosKK12} endpoint which also communicates with a DBpedia endpoint through the use of the \code{SERVICE} keyword in queries. If no query has been generated, the user is notified that the question could not be answered.

%%%%%%%%%%%%%%%%%%%%%%%%%%%%%%%%%%%%%%%%%%%%%%%%%%%%%%%%%%
%%%%%%%%%%%%%%%%%%%%%%%%%%%
%%%%%%%%%%%%%%%%%

\section{Evaluation}\label{sec:evaluation}
The current version of the GeoQA engine presented above has been evaluated using the gold standard dataset GeoData201 and questions GeoQuestions201 presented in Sections~\ref{sec:GeoData} and~\ref{sec:Benchmark}. GeoQA was run using the 153 questions that fall under categories 1 to 6 and only the count questions from category 7 i.e. \qq{How many hospitals are there in Oxford?}. The questions like \qq{What is the longest bridge in Scotland?} fall under category 7 but are not targeted at the moment.

Table \ref{tab:evaluation} summarizes the effectiveness of GeoQA on the gold standard using the well-known metrics of precision, recall and F1. We calculate precision, recall and F1 using the formulas below for the 153 questions. 

$Precision = (|{Correct Answers} \cap {Retrieved Answers}|) / |{Retrived Answers}|$

$Recall = (|{Correct Answers} \cap {Retrieved Answers}|) / |{Correct Answers}|$

$F1 = 2 * ((Precision * Recall) / (Precision + Recall))$

We calculated these f-measures for each questions individually and considered average of them as final f-measure for our system. Table ~\ref{tab:evaluation} presents the average of precision,recall and F1 on all 153 questions. 
For the 22 questions (out of 153), GeoQA does not identify the correct question pattern and as a result no query is generated. From 131 generated queries, only 115 queries return results. The remaining 16 queries do not generate any answers, although such answers exist in the GeoData201 dataset, because of errors in different component that we will discuss further and summerise in Table~\ref{tab:resultbreakdown}. 
%instance identifier
There are some questions for which the instance identifier component fails to annotate the correct entity or does not annotate any entity in the question. E.g., in Q37 \qq{Which counties of Scotland border England?}, TagMe disambiguate to \code{dbr:Scottish\_Borders} that is wrong instance. 
%concept identifier
For some of the questions the concept identifier fails to identify the appropriate class. E.g., in Q146 \qq{Which city council includes Dublin?}, the concept identifier identifies class \textit{city} instead of \textit{city council}. 
%Property Identifier
Sometimes the property identifier component fails to map or to infer the correct property from question text. E.g., in Q86 \qq{Which villages in Scotland have a population of less than 500 people?}, Property identifier maps population to \code{dbp:population} while the most common property for villages in Scotland inside DBpedia is \code{dbp:populationTotal}. In Q80 \qq{Is there a mountain in the county of Greater Manchester taller than 1300 meters above sea level?}, the question is asking about elevation of mountain and property identifier fails to find the right property.
%GeoSPARQL generator
Some other times, the query generator module fails to identify the appropriate pattern for some of the questions. E.g., in Q29 \qq{Which airports are in the city of Salford?}, the identified pattern must be CRI but instead, the query generator identifies CRIRI. Also sometimes the selection of wrong queries would result in no answers. E.g., in Q151 \qq{Which pubs are near Mercure Hotel in Glasgow, Scotland?} the query generator selects a query containing \code{owl:sameAs dbr:Mercure\_Hotels} which is not linked in our dataset resulting in 0 answers.

\begin{table}
\small
\begin{center}
\caption{Evaluation of GeoQA }
\begin{tabular}{ | c | c | c | c | c | }
\hline
\textbf{Gold} & \textbf{Answered} & \textbf {Precision} & \textbf{Recall} & \textbf{F1}\\
\textbf{Questions} & \textbf{Questions} & & &\\
\hline
153 & 131 & 64.63\% & 65.10\% & 62.21\% \\
 \hline
\end{tabular}
\label{tab:evaluation}
\end{center}
\end{table}

\begin{table}
\small
\caption{Breakdown of problematic questions according to pattern they follow}
\begin{tabular}{ | m{0.1\textwidth}| m{0.1\textwidth} | m{0.1\textwidth} | m{0.08\textwidth} | m{0.06\textwidth} | m{0.06\textwidth} | m{0.1\textwidth} | m{0.08\textwidth} |}
\hline
\textbf{} & \multicolumn{7}{ c |}{\textbf{Question Number}}  \\ 
\hline
\textbf{Module Responsible} & \textbf{CRI} & \textbf{CRIRI} & \textbf{CRCRI} & \textbf{IRI} & \textbf{IP} & \textbf{PCRI} & \textbf{PCRIRI}\\
\hline
 Instance Identifier & Q159, Q172 & Q37, Q38, Q63, Q64, Q67, Q182, Q194 & & Q180 & Q40, Q158 & & Q100 \\
 %Q40,Q67,Q100,Q201,Q172,Q37,Q38,Q63,Q108,Q129,Q151
 \hline
 Concept Identifier & Q4, Q118, Q146, Q149 &  & & & & Q177 & \\
 %Q177,Q146,Q48,Q4,Q94,Q118,Q198,Q158,Q38,Q39,Q126,Q159
 \hline
% Relation Identifier & & & & & & &\\
 %Q31,Q146
 %\hline
 Property Identifier &  & & & & & Q80, Q82, Q86 & \\
 %Q80,Q86,Q120
 \hline
 Query Generator & Q23, Q29, Q68, Q92, Q127 & Q13, Q38, Q48, Q77, Q126, Q129, Q151, Q182, Q196 & Q164, Q170 & Q127, Q168, Q180 &  & Q187 & Q120\\ %&& &&&&& \\
 %Q13,Q64,Q22,Q127,Q170,Q182,Q194,Q180,Q161,Q164,Q168,Q92,Q77,Q68,Q23,Q151,Q29,Q38,Q129,Q187
 \hline
\end{tabular}
\label{tab:resultbreakdown}
\end{table}
%%%%%%%%%%%%%%%%%%%%%%%%%%%%%%%%%%%%%%%%%%%%%%%%%%%%%%%%%%%%%%%%%%%%%%%%%%%%%%%%%%%%%%%%%%%%%
%%%%%%%%%%%%%%%%%%%%%%%%%%%%%%%%%%%%%%%%%%%%%%%%%%%%%%%%%%%%%%%%%%%%%%%%%%%%%%%%%%%%%%%%%%%%%

%%%%%%%%%%%%%%%%%%%%%%%%%%%%%%%%%%%%%%%%%%%%%%%%%%%%%%%%%%%%%%%%%%%%%%%%%%%%%%%%%%%%%%%%%%%%%
\section{Conclusion}\label{sec:conclusion}
%%%%%%%%%%%%%%%%%%%%%%%%%%%%%%%%%%%%%%%%%%%%%%%%%%%%%%%%%%%%%%%%%%%%%%%%%%%%%%%%%%%%%%%%%%%%%
%%%%%%%%%%%%%%%%%%%%%%%%%%%%%%%%%%%%%%%%%%%%%%%%%%%%%%%%%%%%%%%%%%%%%%%%%%%%%%%%%%%%%%%%%%%%%
In this paper we have addressed the challenges of providing access to linked geospatial data for non-expert users using natural language QA interfaces. Given the use of geospatial contexts in many practical situations this challenge is of major importance while adopting QA for wide use. Our main contribution was the implementation of GeoQA which is, to the best of our knowledge, the first QA engine which is able to answer questions with a geospatial dimension. We have also evaluated GeoQA using a gold standard dataset and set of questions which we make publicly available so it can also be used by other researchers. 

In future work we plan to improve all the components of GeoQA so we can increase precision, recall and F1 measure further. In addition, we plan to deal with more complicated questions going beyond the question patterns we discussed in this paper. Finally, we plan to consider temporal questions~\cite{DBLP:conf/cikm/JiaARSW18} since temporal and spatial questions arise together naturally in many application contexts.
%%%%%%%%%%%%%%%%%%%%%%%%%%%%%%%%%%%%%%%%%%%%%%%%%%%%%%%%%%%%%%%%%%%%%%%%%%%%%%%%%%%%%%%%%%%%%
%%%%%%%%%%%%%%%%%%%%%%%%%%%%%%%%%%%%%%%%%%%%%%%%%%%%%%%%%%%%%%%%%%%%%%%%%%%%%%%%%%%%%%%%%%%%% 

\textbf{Funding:} This work has been funded by the Horizon 2020 Marie Sklodowska-Curie Innovative Training Network WDAqua (Answering Questions using Web Data) with grant agreement No. 642795 and by Hellenic Foundation for Research and Innovation (HFRI) and the General Secretariat for Research and Technology (GSRT), under the grant GeoQa (GA. no HFRI-FM17-2351).



%% References



%% New version of the num-names style
\bibliographystyle{elsarticle-num-names}
\bibliography{geoqa.bib}

\begin{thebibliography}{47}
\expandafter\ifx\csname natexlab\endcsname\relax\def\natexlab#1{#1}\fi
\providecommand{\url}[1]{\texttt{#1}}
\providecommand{\href}[2]{#2}
\providecommand{\path}[1]{#1}
\providecommand{\DOIprefix}{doi:}
\providecommand{\ArXivprefix}{arXiv:}
\providecommand{\URLprefix}{URL: }
\providecommand{\Pubmedprefix}{pmid:}
\providecommand{\doi}[1]{\href{http://dx.doi.org/#1}{\path{#1}}}
\providecommand{\Pubmed}[1]{\href{pmid:#1}{\path{#1}}}
\providecommand{\bibinfo}[2]{#2}
\ifx\xfnm\relax \def\xfnm[#1]{\unskip,\space#1}\fi
%Type = Article
\bibitem[{Hirschman and Gaizauskas(2001)}]{DBLP:journals/nle/HirschmanG01}
\bibinfo{author}{L.~Hirschman}, \bibinfo{author}{R.~J. Gaizauskas},
\newblock \bibinfo{title}{Natural language question answering: the view from
  here},
\newblock \bibinfo{journal}{Nat. Lang. Eng.} \bibinfo{volume}{7}
  (\bibinfo{year}{2001}) \bibinfo{pages}{275--300}. \URLprefix
  \url{https://doi.org/10.1017/S1351324901002807}.
  \DOIprefix\doi{10.1017/S1351324901002807}.
%Type = Article
\bibitem[{Lee and Kang(2015)}]{DBLP:journals/bdr/LeeK15}
\bibinfo{author}{J.~Lee}, \bibinfo{author}{M.~Kang},
\newblock \bibinfo{title}{Geospatial big data: Challenges and opportunities},
\newblock \bibinfo{journal}{Big Data Res.} \bibinfo{volume}{2}
  (\bibinfo{year}{2015}) \bibinfo{pages}{74--81}. \URLprefix
  \url{https://doi.org/10.1016/j.bdr.2015.01.003}.
  \DOIprefix\doi{10.1016/j.bdr.2015.01.003}.
%Type = Inproceedings
\bibitem[{Auer et~al.(2009)Auer, Lehmann, and
  Hellmann}]{DBLP:conf/semweb/AuerLH09}
\bibinfo{author}{S.~Auer}, \bibinfo{author}{J.~Lehmann},
  \bibinfo{author}{S.~Hellmann},
\newblock \bibinfo{title}{Linkedgeodata: Adding a spatial dimension to the web
  of data},
\newblock in: \bibinfo{editor}{A.~Bernstein}, \bibinfo{editor}{D.~R. Karger},
  \bibinfo{editor}{T.~Heath}, \bibinfo{editor}{L.~Feigenbaum},
  \bibinfo{editor}{D.~Maynard}, \bibinfo{editor}{E.~Motta},
  \bibinfo{editor}{K.~Thirunarayan} (Eds.), \bibinfo{booktitle}{The Semantic
  Web - {ISWC} 2009, 8th International Semantic Web Conference, {ISWC} 2009,
  Chantilly, VA, USA, October 25-29, 2009. Proceedings}, volume
  \bibinfo{volume}{5823} of \textit{\bibinfo{series}{Lecture Notes in Computer
  Science}}, \bibinfo{publisher}{Springer}, \bibinfo{year}{2009}, pp.
  \bibinfo{pages}{731--746}. \URLprefix
  \url{https://doi.org/10.1007/978-3-642-04930-9\_46}.
  \DOIprefix\doi{10.1007/978-3-642-04930-9\_46}.
%Type = Inproceedings
\bibitem[{Kyzirakos et~al.(2012)Kyzirakos, Karpathiotakis, and
  Koubarakis}]{DBLP:conf/semweb/KyzirakosKK12}
\bibinfo{author}{K.~Kyzirakos}, \bibinfo{author}{M.~Karpathiotakis},
  \bibinfo{author}{M.~Koubarakis},
\newblock \bibinfo{title}{Strabon: {A} semantic geospatial {DBMS}},
\newblock in: \bibinfo{editor}{P.~Cudr{\'{e}}{-}Mauroux},
  \bibinfo{editor}{J.~Heflin}, \bibinfo{editor}{E.~Sirin},
  \bibinfo{editor}{T.~Tudorache}, \bibinfo{editor}{J.~Euzenat},
  \bibinfo{editor}{M.~Hauswirth}, \bibinfo{editor}{J.~X. Parreira},
  \bibinfo{editor}{J.~Hendler}, \bibinfo{editor}{G.~Schreiber},
  \bibinfo{editor}{A.~Bernstein}, \bibinfo{editor}{E.~Blomqvist} (Eds.),
  \bibinfo{booktitle}{The Semantic Web - {ISWC} 2012 - 11th International
  Semantic Web Conference, Boston, MA, USA, November 11-15, 2012, Proceedings,
  Part {I}}, volume \bibinfo{volume}{7649} of \textit{\bibinfo{series}{Lecture
  Notes in Computer Science}}, \bibinfo{publisher}{Springer},
  \bibinfo{year}{2012}, pp. \bibinfo{pages}{295--311}. \URLprefix
  \url{https://doi.org/10.1007/978-3-642-35176-1\_19}.
  \DOIprefix\doi{10.1007/978-3-642-35176-1\_19}.
%Type = Article
\bibitem[{Diefenbach et~al.(2018)Diefenbach, L{\'{o}}pez, Singh, and
  Maret}]{DBLP:journals/kais/DiefenbachLSM18}
\bibinfo{author}{D.~Diefenbach}, \bibinfo{author}{V.~L{\'{o}}pez},
  \bibinfo{author}{K.~D. Singh}, \bibinfo{author}{P.~Maret},
\newblock \bibinfo{title}{Core techniques of question answering systems over
  knowledge bases: a survey},
\newblock \bibinfo{journal}{Knowl. Inf. Syst.} \bibinfo{volume}{55}
  (\bibinfo{year}{2018}) \bibinfo{pages}{529--569}. \URLprefix
  \url{https://doi.org/10.1007/s10115-017-1100-y}.
  \DOIprefix\doi{10.1007/s10115-017-1100-y}.
%Type = Inproceedings
\bibitem[{Schlaisich and Egenhofer(2001)}]{DBLP:conf/hci/SchlaisichE01}
\bibinfo{author}{I.~Schlaisich}, \bibinfo{author}{M.~J. Egenhofer},
\newblock \bibinfo{title}{Multimodal spatial querying: what people sketch and
  talk about},
\newblock in: \bibinfo{editor}{C.~Stephanidis} (Ed.),
  \bibinfo{booktitle}{Universal Access In {HCI:} Towards an Information Society
  for All, Proceedings of {HCI} International '2001 (the 9th International
  Conference on Human-Computer Interaction), New Orleans, USA, August 5-10,
  2001, Volume 3}, \bibinfo{publisher}{Lawrence Erlbaum}, \bibinfo{year}{2001},
  pp. \bibinfo{pages}{732--736}.
%Type = Inproceedings
\bibitem[{Singh et~al.(2016)Singh, Both, Diefenbach, Shekarpour, Cherix, and
  Lange}]{DBLP:conf/esws/SinghBDSC016}
\bibinfo{author}{K.~Singh}, \bibinfo{author}{A.~Both},
  \bibinfo{author}{D.~Diefenbach}, \bibinfo{author}{S.~Shekarpour},
  \bibinfo{author}{D.~Cherix}, \bibinfo{author}{C.~Lange},
\newblock \bibinfo{title}{Qanary - the fast track to creating a question
  answering system with linked data technology},
\newblock in: \bibinfo{editor}{H.~Sack}, \bibinfo{editor}{G.~Rizzo},
  \bibinfo{editor}{N.~Steinmetz}, \bibinfo{editor}{D.~Mladenic},
  \bibinfo{editor}{S.~Auer}, \bibinfo{editor}{C.~Lange} (Eds.),
  \bibinfo{booktitle}{The Semantic Web - {ESWC} 2016 Satellite Events,
  Heraklion, Crete, Greece, May 29 - June 2, 2016, Revised Selected Papers},
  volume \bibinfo{volume}{9989} of \textit{\bibinfo{series}{Lecture Notes in
  Computer Science}}, \bibinfo{year}{2016}, pp. \bibinfo{pages}{183--188}.
  \URLprefix \url{https://doi.org/10.1007/978-3-319-47602-5\_36}.
  \DOIprefix\doi{10.1007/978-3-319-47602-5\_36}.
%Type = Inproceedings
\bibitem[{Both et~al.(2016)Both, Diefenbach, Singh, Shekarpour, Cherix, and
  Lange}]{DBLP:conf/esws/BothDSSC016}
\bibinfo{author}{A.~Both}, \bibinfo{author}{D.~Diefenbach},
  \bibinfo{author}{K.~Singh}, \bibinfo{author}{S.~Shekarpour},
  \bibinfo{author}{D.~Cherix}, \bibinfo{author}{C.~Lange},
\newblock \bibinfo{title}{Qanary - {A} methodology for vocabulary-driven open
  question answering systems},
\newblock in: \bibinfo{editor}{H.~Sack}, \bibinfo{editor}{E.~Blomqvist},
  \bibinfo{editor}{M.~d'Aquin}, \bibinfo{editor}{C.~Ghidini},
  \bibinfo{editor}{S.~P. Ponzetto}, \bibinfo{editor}{C.~Lange} (Eds.),
  \bibinfo{booktitle}{The Semantic Web. Latest Advances and New Domains - 13th
  International Conference, {ESWC} 2016, Heraklion, Crete, Greece, May 29 -
  June 2, 2016, Proceedings}, volume \bibinfo{volume}{9678} of
  \textit{\bibinfo{series}{Lecture Notes in Computer Science}},
  \bibinfo{publisher}{Springer}, \bibinfo{year}{2016}, pp.
  \bibinfo{pages}{625--641}. \URLprefix
  \url{https://doi.org/10.1007/978-3-319-34129-3\_38}.
  \DOIprefix\doi{10.1007/978-3-319-34129-3\_38}.
%Type = Inproceedings
\bibitem[{Singh et~al.(2018)Singh, Radhakrishna, Both, Shekarpour, Lytra,
  Usbeck, Vyas, Khikmatullaev, Punjani, Lange, Vidal, Lehmann, and
  Auer}]{DBLP:conf/www/SinghRBSLUVKP0V18}
\bibinfo{author}{K.~Singh}, \bibinfo{author}{A.~S. Radhakrishna},
  \bibinfo{author}{A.~Both}, \bibinfo{author}{S.~Shekarpour},
  \bibinfo{author}{I.~Lytra}, \bibinfo{author}{R.~Usbeck},
  \bibinfo{author}{A.~Vyas}, \bibinfo{author}{A.~Khikmatullaev},
  \bibinfo{author}{D.~Punjani}, \bibinfo{author}{C.~Lange},
  \bibinfo{author}{M.~Vidal}, \bibinfo{author}{J.~Lehmann},
  \bibinfo{author}{S.~Auer},
\newblock \bibinfo{title}{Why reinvent the wheel: Let's build question
  answering systems together},
\newblock in: \bibinfo{editor}{P.~Champin}, \bibinfo{editor}{F.~L. Gandon},
  \bibinfo{editor}{M.~Lalmas}, \bibinfo{editor}{P.~G. Ipeirotis} (Eds.),
  \bibinfo{booktitle}{Proceedings of the 2018 World Wide Web Conference on
  World Wide Web, {WWW} 2018, Lyon, France, April 23-27, 2018},
  \bibinfo{publisher}{{ACM}}, \bibinfo{year}{2018}, pp.
  \bibinfo{pages}{1247--1256}. \URLprefix
  \url{https://doi.org/10.1145/3178876.3186023}.
  \DOIprefix\doi{10.1145/3178876.3186023}.
%Type = Inproceedings
\bibitem[{Punjani et~al.(2018)Punjani, Singh, Both, Koubarakis, Angelidis,
  Bereta, Beris, Bilidas, Ioannidis, Karalis et~al.}]{punjani2018template}
\bibinfo{author}{D.~Punjani}, \bibinfo{author}{K.~Singh},
  \bibinfo{author}{A.~Both}, \bibinfo{author}{M.~Koubarakis},
  \bibinfo{author}{I.~Angelidis}, \bibinfo{author}{K.~Bereta},
  \bibinfo{author}{T.~Beris}, \bibinfo{author}{D.~Bilidas},
  \bibinfo{author}{T.~Ioannidis}, \bibinfo{author}{N.~Karalis}, et~al.,
\newblock \bibinfo{title}{Template-based question answering over linked
  geospatial data},
\newblock in: \bibinfo{booktitle}{Proceedings of the 12th Workshop on
  Geographic Information Retrieval}, \bibinfo{year}{2018}, pp.
  \bibinfo{pages}{1--10}.
%Type = Inproceedings
\bibitem[{Jones et~al.(2002)Jones, Purves, Ruas, Sanderson, Sester, van
  Kreveld, and Weibel}]{DBLP:conf/sigir/JonesPRSSKW02}
\bibinfo{author}{C.~B. Jones}, \bibinfo{author}{R.~Purves},
  \bibinfo{author}{A.~Ruas}, \bibinfo{author}{M.~Sanderson},
  \bibinfo{author}{M.~Sester}, \bibinfo{author}{M.~J. van Kreveld},
  \bibinfo{author}{R.~Weibel},
\newblock \bibinfo{title}{Spatial information retrieval and geographical
  ontologies an overview of the {SPIRIT} project},
\newblock in: \bibinfo{editor}{K.~J{\"{a}}rvelin},
  \bibinfo{editor}{M.~Beaulieu}, \bibinfo{editor}{R.~A. Baeza{-}Yates},
  \bibinfo{editor}{S.~Myaeng} (Eds.), \bibinfo{booktitle}{{SIGIR} 2002:
  Proceedings of the 25th Annual International {ACM} {SIGIR} Conference on
  Research and Development in Information Retrieval, August 11-15, 2002,
  Tampere, Finland}, \bibinfo{publisher}{{ACM}}, \bibinfo{year}{2002}, pp.
  \bibinfo{pages}{387--388}. \URLprefix
  \url{https://doi.org/10.1145/564376.564457}.
  \DOIprefix\doi{10.1145/564376.564457}.
%Type = Inproceedings
\bibitem[{Lieberman et~al.(2007)Lieberman, Samet, Sankaranarayanan, and
  Sperling}]{DBLP:conf/gis/LiebermanSSS07}
\bibinfo{author}{M.~D. Lieberman}, \bibinfo{author}{H.~Samet},
  \bibinfo{author}{J.~Sankaranarayanan}, \bibinfo{author}{J.~Sperling},
\newblock \bibinfo{title}{{STEWARD:} architecture of a spatio-textual search
  engine},
\newblock in: \bibinfo{editor}{H.~Samet}, \bibinfo{editor}{C.~Shahabi},
  \bibinfo{editor}{M.~Schneider} (Eds.), \bibinfo{booktitle}{15th {ACM}
  International Symposium on Geographic Information Systems, {ACM-GIS} 2007,
  November 7-9, 2007, Seattle, Washington, USA, Proceedings},
  \bibinfo{publisher}{{ACM}}, \bibinfo{year}{2007}, p.~\bibinfo{pages}{25}.
  \URLprefix \url{https://doi.org/10.1145/1341012.1341045}.
  \DOIprefix\doi{10.1145/1341012.1341045}.
%Type = Inproceedings
\bibitem[{Santos et~al.(2008)Santos, Cardoso, Carvalho, Dornescu, Hartrumpf,
  Leveling, and Skalban}]{Santos2008GettingGA}
\bibinfo{author}{D.~Santos}, \bibinfo{author}{N.~Cardoso},
  \bibinfo{author}{P.~Carvalho}, \bibinfo{author}{I.~Dornescu},
  \bibinfo{author}{S.~Hartrumpf}, \bibinfo{author}{J.~Leveling},
  \bibinfo{author}{Y.~Skalban},
\newblock \bibinfo{title}{Getting geographical answers from wikipedia: the
  gikip pilot at clef},
\newblock in: \bibinfo{booktitle}{CLEF}, \bibinfo{year}{2008}.
%Type = Inproceedings
\bibitem[{Bereta and Koubarakis(2016)}]{DBLP:conf/semweb/BeretaK16}
\bibinfo{author}{K.~Bereta}, \bibinfo{author}{M.~Koubarakis},
\newblock \bibinfo{title}{Ontop of geospatial databases},
\newblock in: \bibinfo{editor}{P.~T. Groth}, \bibinfo{editor}{E.~Simperl},
  \bibinfo{editor}{A.~J.~G. Gray}, \bibinfo{editor}{M.~Sabou},
  \bibinfo{editor}{M.~Kr{\"{o}}tzsch}, \bibinfo{editor}{F.~L{\'{e}}cu{\'{e}}},
  \bibinfo{editor}{F.~Fl{\"{o}}ck}, \bibinfo{editor}{Y.~Gil} (Eds.),
  \bibinfo{booktitle}{The Semantic Web - {ISWC} 2016 - 15th International
  Semantic Web Conference, Kobe, Japan, October 17-21, 2016, Proceedings, Part
  {I}}, volume \bibinfo{volume}{9981} of \textit{\bibinfo{series}{Lecture Notes
  in Computer Science}}, \bibinfo{year}{2016}, pp. \bibinfo{pages}{37--52}.
  \URLprefix \url{https://doi.org/10.1007/978-3-319-46523-4\_3}.
  \DOIprefix\doi{10.1007/978-3-319-46523-4\_3}.
%Type = Inproceedings
\bibitem[{Younis et~al.(2012)Younis, Jones, Tanasescu, and
  Abdelmoty}]{DBLP:conf/giscience/YounisJTA12}
\bibinfo{author}{E.~M.~G. Younis}, \bibinfo{author}{C.~B. Jones},
  \bibinfo{author}{V.~Tanasescu}, \bibinfo{author}{A.~I. Abdelmoty},
\newblock \bibinfo{title}{Hybrid geo-spatial query methods on the semantic web
  with a spatially-enhanced index of dbpedia},
\newblock in: \bibinfo{editor}{N.~Xiao}, \bibinfo{editor}{M.~Kwan},
  \bibinfo{editor}{M.~F. Goodchild}, \bibinfo{editor}{S.~Shekhar} (Eds.),
  \bibinfo{booktitle}{Geographic Information Science - 7th International
  Conference, GIScience 2012, Columbus, OH, USA, September 18-21, 2012.
  Proceedings}, volume \bibinfo{volume}{7478} of
  \textit{\bibinfo{series}{Lecture Notes in Computer Science}},
  \bibinfo{publisher}{Springer}, \bibinfo{year}{2012}, pp.
  \bibinfo{pages}{340--353}. \URLprefix
  \url{https://doi.org/10.1007/978-3-642-33024-7\_25}.
  \DOIprefix\doi{10.1007/978-3-642-33024-7\_25}.
%Type = Article
\bibitem[{Gr{\"{u}}tter et~al.(2017)Gr{\"{u}}tter, Purves, and
  Wotruba}]{DBLP:journals/tgis/GrutterPW17}
\bibinfo{author}{R.~Gr{\"{u}}tter}, \bibinfo{author}{R.~S. Purves},
  \bibinfo{author}{L.~Wotruba},
\newblock \bibinfo{title}{Evaluating topological queries in linked data using
  dbpedia and geonames in switzerland and scotland},
\newblock \bibinfo{journal}{Trans. {GIS}} \bibinfo{volume}{21}
  (\bibinfo{year}{2017}) \bibinfo{pages}{114--133}. \URLprefix
  \url{https://doi.org/10.1111/tgis.12196}. \DOIprefix\doi{10.1111/tgis.12196}.
%Type = Inproceedings
\bibitem[{Hamzei et~al.(2019)Hamzei, Li, Vasardani, Baldwin, Winter, and
  Tomko}]{DBLP:conf/agile/HamzeiLVB0T19}
\bibinfo{author}{E.~Hamzei}, \bibinfo{author}{H.~Li},
  \bibinfo{author}{M.~Vasardani}, \bibinfo{author}{T.~Baldwin},
  \bibinfo{author}{S.~Winter}, \bibinfo{author}{M.~Tomko},
\newblock \bibinfo{title}{Place questions and human-generated answers: {A} data
  analysis approach},
\newblock in: \bibinfo{editor}{P.~C. Kyriakidis}, \bibinfo{editor}{D.~G.
  Hadjimitsis}, \bibinfo{editor}{D.~Skarlatos}, \bibinfo{editor}{A.~Mansourian}
  (Eds.), \bibinfo{booktitle}{Geospatial Technologies for Local and Regional
  Development - Proceedings of the 22nd {AGILE} Conference on Geographic
  Information Science, Limassol, Cyprus, June 17-20, 2019}, Lecture Notes in
  Geoinformation and Cartography, \bibinfo{publisher}{Springer},
  \bibinfo{year}{2019}, pp. \bibinfo{pages}{3--19}. \URLprefix
  \url{https://doi.org/10.1007/978-3-030-14745-7\_1}.
  \DOIprefix\doi{10.1007/978-3-030-14745-7\_1}.
%Type = Inproceedings
\bibitem[{Nguyen et~al.(2016)Nguyen, Rosenberg, Song, Gao, Tiwary, Majumder,
  and Deng}]{DBLP:conf/nips/NguyenRSGTMD16}
\bibinfo{author}{T.~Nguyen}, \bibinfo{author}{M.~Rosenberg},
  \bibinfo{author}{X.~Song}, \bibinfo{author}{J.~Gao},
  \bibinfo{author}{S.~Tiwary}, \bibinfo{author}{R.~Majumder},
  \bibinfo{author}{L.~Deng},
\newblock \bibinfo{title}{{MS} {MARCO:} {A} human generated machine reading
  comprehension dataset},
\newblock in: \bibinfo{editor}{T.~R. Besold}, \bibinfo{editor}{A.~Bordes},
  \bibinfo{editor}{A.~S. d'Avila Garcez}, \bibinfo{editor}{G.~Wayne} (Eds.),
  \bibinfo{booktitle}{Proceedings of the Workshop on Cognitive Computation:
  Integrating neural and symbolic approaches 2016 co-located with the 30th
  Annual Conference on Neural Information Processing Systems {(NIPS} 2016),
  Barcelona, Spain, December 9, 2016}, volume \bibinfo{volume}{1773} of
  \textit{\bibinfo{series}{{CEUR} Workshop Proceedings}},
  \bibinfo{publisher}{CEUR-WS.org}, \bibinfo{year}{2016}.
%Type = Inproceedings
\bibitem[{Tang and Mooney(2001)}]{DBLP:conf/ecml/TangM01}
\bibinfo{author}{L.~R. Tang}, \bibinfo{author}{R.~J. Mooney},
\newblock \bibinfo{title}{Using multiple clause constructors in inductive logic
  programming for semantic parsing},
\newblock in: \bibinfo{editor}{L.~D. Raedt}, \bibinfo{editor}{P.~A. Flach}
  (Eds.), \bibinfo{booktitle}{Machine Learning: {EMCL} 2001, 12th European
  Conference on Machine Learning, Freiburg, Germany, September 5-7, 2001,
  Proceedings}, volume \bibinfo{volume}{2167} of
  \textit{\bibinfo{series}{Lecture Notes in Computer Science}},
  \bibinfo{publisher}{Springer}, \bibinfo{year}{2001}, pp.
  \bibinfo{pages}{466--477}. \URLprefix
  \url{https://doi.org/10.1007/3-540-44795-4\_40}.
  \DOIprefix\doi{10.1007/3-540-44795-4\_40}.
%Type = Article
\bibitem[{L{\'{o}}pez et~al.(2012)L{\'{o}}pez, Fern{\'{a}}ndez, Motta, and
  Stieler}]{DBLP:journals/semweb/LopezFMS12}
\bibinfo{author}{V.~L{\'{o}}pez}, \bibinfo{author}{M.~Fern{\'{a}}ndez},
  \bibinfo{author}{E.~Motta}, \bibinfo{author}{N.~Stieler},
\newblock \bibinfo{title}{Poweraqua: Supporting users in querying and exploring
  the semantic web},
\newblock \bibinfo{journal}{Semantic Web} \bibinfo{volume}{3}
  (\bibinfo{year}{2012}) \bibinfo{pages}{249--265}. \URLprefix
  \url{https://doi.org/10.3233/SW-2011-0030}.
  \DOIprefix\doi{10.3233/SW-2011-0030}.
%Type = Inproceedings
\bibitem[{Salas and Harth(2011)}]{salas2011finding}
\bibinfo{author}{J.~Salas}, \bibinfo{author}{A.~Harth},
\newblock \bibinfo{title}{Finding spatial equivalences accross multiple rdf
  datasets},
\newblock in: \bibinfo{booktitle}{Proceedings of the Terra Cognita Workshop on
  Foundations, Technologies and Applications of the Geospatial Web},
  \bibinfo{organization}{Citeseer}, \bibinfo{year}{2011}, pp.
  \bibinfo{pages}{114--126}.
%Type = Article
\bibitem[{Stadler et~al.(2012)Stadler, Lehmann, H{\"{o}}ffner, and
  Auer}]{DBLP:journals/semweb/StadlerLHA12}
\bibinfo{author}{C.~Stadler}, \bibinfo{author}{J.~Lehmann},
  \bibinfo{author}{K.~H{\"{o}}ffner}, \bibinfo{author}{S.~Auer},
\newblock \bibinfo{title}{Linkedgeodata: {A} core for a web of spatial open
  data},
\newblock \bibinfo{journal}{Semantic Web} \bibinfo{volume}{3}
  (\bibinfo{year}{2012}) \bibinfo{pages}{333--354}. \URLprefix
  \url{https://doi.org/10.3233/SW-2011-0052}.
  \DOIprefix\doi{10.3233/SW-2011-0052}.
%Type = Inproceedings
\bibitem[{Volz et~al.(2009)Volz, Bizer, Gaedke, and
  Kobilarov}]{DBLP:conf/www/VolzBGK09}
\bibinfo{author}{J.~Volz}, \bibinfo{author}{C.~Bizer},
  \bibinfo{author}{M.~Gaedke}, \bibinfo{author}{G.~Kobilarov},
\newblock \bibinfo{title}{Silk - {A} link discovery framework for the web of
  data},
\newblock in: \bibinfo{editor}{C.~Bizer}, \bibinfo{editor}{T.~Heath},
  \bibinfo{editor}{T.~Berners{-}Lee}, \bibinfo{editor}{K.~Idehen} (Eds.),
  \bibinfo{booktitle}{Proceedings of the {WWW2009} Workshop on Linked Data on
  the Web, {LDOW} 2009, Madrid, Spain, April 20, 2009}, volume
  \bibinfo{volume}{538} of \textit{\bibinfo{series}{{CEUR} Workshop
  Proceedings}}, \bibinfo{publisher}{CEUR-WS.org}, \bibinfo{year}{2009}.
%Type = Inproceedings
\bibitem[{Ho et~al.(2019)Ho, Ibrahim, Pal, Berberich, and
  Weikum}]{DBLP:conf/semweb/HoIPBW19}
\bibinfo{author}{V.~T. Ho}, \bibinfo{author}{Y.~Ibrahim},
  \bibinfo{author}{K.~Pal}, \bibinfo{author}{K.~Berberich},
  \bibinfo{author}{G.~Weikum},
\newblock \bibinfo{title}{Qsearch: Answering quantity queries from text},
\newblock in: \bibinfo{booktitle}{The Semantic Web - {ISWC} 2019 - 18th
  International Semantic Web Conference, Auckland, New Zealand, October 26-30,
  2019, Proceedings, Part {I}}, \bibinfo{year}{2019}, pp.
  \bibinfo{pages}{237--257}.
%Type = Inproceedings
\bibitem[{Diefenbach et~al.(2017)Diefenbach, Singh, Both, Cherix, Lange, and
  Auer}]{DBLP:conf/icwe/DiefenbachSBC0A17}
\bibinfo{author}{D.~Diefenbach}, \bibinfo{author}{K.~Singh},
  \bibinfo{author}{A.~Both}, \bibinfo{author}{D.~Cherix},
  \bibinfo{author}{C.~Lange}, \bibinfo{author}{S.~Auer},
\newblock \bibinfo{title}{The qanary ecosystem: Getting new insights by
  composing question answering pipelines},
\newblock in: \bibinfo{editor}{J.~Cabot}, \bibinfo{editor}{R.~D. Virgilio},
  \bibinfo{editor}{R.~Torlone} (Eds.), \bibinfo{booktitle}{Web Engineering -
  17th International Conference, {ICWE} 2017, Rome, Italy, June 5-8, 2017,
  Proceedings}, volume \bibinfo{volume}{10360} of
  \textit{\bibinfo{series}{Lecture Notes in Computer Science}},
  \bibinfo{publisher}{Springer}, \bibinfo{year}{2017}, pp.
  \bibinfo{pages}{171--189}. \URLprefix
  \url{https://doi.org/10.1007/978-3-319-60131-1\_10}.
  \DOIprefix\doi{10.1007/978-3-319-60131-1\_10}.
%Type = Inproceedings
\bibitem[{Both et~al.(2017)Both, Singh, Diefenbach, and
  Lytra}]{DBLP:conf/icwe/BothSDL17}
\bibinfo{author}{A.~Both}, \bibinfo{author}{K.~Singh},
  \bibinfo{author}{D.~Diefenbach}, \bibinfo{author}{I.~Lytra},
\newblock \bibinfo{title}{Rapid engineering of {QA} systems using the
  light-weight qanary architecture},
\newblock in: \bibinfo{editor}{J.~Cabot}, \bibinfo{editor}{R.~D. Virgilio},
  \bibinfo{editor}{R.~Torlone} (Eds.), \bibinfo{booktitle}{Web Engineering -
  17th International Conference, {ICWE} 2017, Rome, Italy, June 5-8, 2017,
  Proceedings}, volume \bibinfo{volume}{10360} of
  \textit{\bibinfo{series}{Lecture Notes in Computer Science}},
  \bibinfo{publisher}{Springer}, \bibinfo{year}{2017}, pp.
  \bibinfo{pages}{544--548}. \URLprefix
  \url{https://doi.org/10.1007/978-3-319-60131-1\_40}.
  \DOIprefix\doi{10.1007/978-3-319-60131-1\_40}.
%Type = Inproceedings
\bibitem[{Dubey et~al.(2016)Dubey, Dasgupta, Sharma, H{\"{o}}ffner, and
  Lehmann}]{DBLP:conf/esws/DubeyDSHL16}
\bibinfo{author}{M.~Dubey}, \bibinfo{author}{S.~Dasgupta},
  \bibinfo{author}{A.~Sharma}, \bibinfo{author}{K.~H{\"{o}}ffner},
  \bibinfo{author}{J.~Lehmann},
\newblock \bibinfo{title}{Asknow: {A} framework for natural language query
  formalization in {SPARQL}},
\newblock in: \bibinfo{editor}{H.~Sack}, \bibinfo{editor}{E.~Blomqvist},
  \bibinfo{editor}{M.~d'Aquin}, \bibinfo{editor}{C.~Ghidini},
  \bibinfo{editor}{S.~P. Ponzetto}, \bibinfo{editor}{C.~Lange} (Eds.),
  \bibinfo{booktitle}{The Semantic Web. Latest Advances and New Domains - 13th
  International Conference, {ESWC} 2016, Heraklion, Crete, Greece, May 29 -
  June 2, 2016, Proceedings}, volume \bibinfo{volume}{9678} of
  \textit{\bibinfo{series}{Lecture Notes in Computer Science}},
  \bibinfo{publisher}{Springer}, \bibinfo{year}{2016}, pp.
  \bibinfo{pages}{300--316}. \URLprefix
  \url{https://doi.org/10.1007/978-3-319-34129-3\_19}.
  \DOIprefix\doi{10.1007/978-3-319-34129-3\_19}.
%Type = Inproceedings
\bibitem[{L{\'{o}}pez et~al.(2006)L{\'{o}}pez, Motta, and
  Uren}]{DBLP:conf/esws/LopezMU06}
\bibinfo{author}{V.~L{\'{o}}pez}, \bibinfo{author}{E.~Motta},
  \bibinfo{author}{V.~S. Uren},
\newblock \bibinfo{title}{Poweraqua: Fishing the semantic web},
\newblock in: \bibinfo{editor}{Y.~Sure}, \bibinfo{editor}{J.~Domingue} (Eds.),
  \bibinfo{booktitle}{The Semantic Web: Research and Applications, 3rd European
  Semantic Web Conference, {ESWC} 2006, Budva, Montenegro, June 11-14, 2006,
  Proceedings}, volume \bibinfo{volume}{4011} of
  \textit{\bibinfo{series}{Lecture Notes in Computer Science}},
  \bibinfo{publisher}{Springer}, \bibinfo{year}{2006}, pp.
  \bibinfo{pages}{393--410}. \URLprefix
  \url{https://doi.org/10.1007/11762256\_30}.
  \DOIprefix\doi{10.1007/11762256\_30}.
%Type = Inproceedings
\bibitem[{Unger et~al.(2012)Unger, B{\"{u}}hmann, Lehmann, Ngomo, Gerber, and
  Cimiano}]{DBLP:conf/www/UngerBLNGC12}
\bibinfo{author}{C.~Unger}, \bibinfo{author}{L.~B{\"{u}}hmann},
  \bibinfo{author}{J.~Lehmann}, \bibinfo{author}{A.~N. Ngomo},
  \bibinfo{author}{D.~Gerber}, \bibinfo{author}{P.~Cimiano},
\newblock \bibinfo{title}{Template-based question answering over {RDF} data},
\newblock in: \bibinfo{editor}{A.~Mille}, \bibinfo{editor}{F.~L. Gandon},
  \bibinfo{editor}{J.~Misselis}, \bibinfo{editor}{M.~Rabinovich},
  \bibinfo{editor}{S.~Staab} (Eds.), \bibinfo{booktitle}{Proceedings of the
  21st World Wide Web Conference 2012, {WWW} 2012, Lyon, France, April 16-20,
  2012}, \bibinfo{publisher}{{ACM}}, \bibinfo{year}{2012}, pp.
  \bibinfo{pages}{639--648}. \URLprefix
  \url{https://doi.org/10.1145/2187836.2187923}.
  \DOIprefix\doi{10.1145/2187836.2187923}.
%Type = Inproceedings
\bibitem[{Nivre et~al.(2016)Nivre, de~Marneffe, Ginter, Goldberg, Hajic,
  Manning, McDonald, Petrov, Pyysalo, Silveira, Tsarfaty, and
  Zeman}]{DBLP:conf/lrec/NivreMGGHMMPPST16}
\bibinfo{author}{J.~Nivre}, \bibinfo{author}{M.~de~Marneffe},
  \bibinfo{author}{F.~Ginter}, \bibinfo{author}{Y.~Goldberg},
  \bibinfo{author}{J.~Hajic}, \bibinfo{author}{C.~D. Manning},
  \bibinfo{author}{R.~T. McDonald}, \bibinfo{author}{S.~Petrov},
  \bibinfo{author}{S.~Pyysalo}, \bibinfo{author}{N.~Silveira},
  \bibinfo{author}{R.~Tsarfaty}, \bibinfo{author}{D.~Zeman},
\newblock \bibinfo{title}{Universal dependencies v1: {A} multilingual treebank
  collection},
\newblock in: \bibinfo{editor}{N.~Calzolari}, \bibinfo{editor}{K.~Choukri},
  \bibinfo{editor}{T.~Declerck}, \bibinfo{editor}{S.~Goggi},
  \bibinfo{editor}{M.~Grobelnik}, \bibinfo{editor}{B.~Maegaard},
  \bibinfo{editor}{J.~Mariani}, \bibinfo{editor}{H.~Mazo},
  \bibinfo{editor}{A.~Moreno}, \bibinfo{editor}{J.~Odijk},
  \bibinfo{editor}{S.~Piperidis} (Eds.), \bibinfo{booktitle}{Proceedings of the
  Tenth International Conference on Language Resources and Evaluation {LREC}
  2016, Portoro{\v{z}}, Slovenia, May 23-28, 2016},
  \bibinfo{publisher}{European Language Resources Association {(ELRA)}},
  \bibinfo{year}{2016}.
%Type = Inproceedings
\bibitem[{Ferragina and Scaiella(2010)}]{DBLP:conf/cikm/FerraginaS10}
\bibinfo{author}{P.~Ferragina}, \bibinfo{author}{U.~Scaiella},
\newblock \bibinfo{title}{{TAGME:} on-the-fly annotation of short text
  fragments (by wikipedia entities)},
\newblock in: \bibinfo{editor}{J.~Huang}, \bibinfo{editor}{N.~Koudas},
  \bibinfo{editor}{G.~J.~F. Jones}, \bibinfo{editor}{X.~Wu},
  \bibinfo{editor}{K.~Collins{-}Thompson}, \bibinfo{editor}{A.~An} (Eds.),
  \bibinfo{booktitle}{Proceedings of the 19th {ACM} Conference on Information
  and Knowledge Management, {CIKM} 2010, Toronto, Ontario, Canada, October
  26-30, 2010}, \bibinfo{publisher}{{ACM}}, \bibinfo{year}{2010}, pp.
  \bibinfo{pages}{1625--1628}. \URLprefix
  \url{https://doi.org/10.1145/1871437.1871689}.
  \DOIprefix\doi{10.1145/1871437.1871689}.
%Type = Inproceedings
\bibitem[{Finkel et~al.(2005)Finkel, Grenager, and
  Manning}]{DBLP:conf/acl/FinkelGM05}
\bibinfo{author}{J.~R. Finkel}, \bibinfo{author}{T.~Grenager},
  \bibinfo{author}{C.~D. Manning},
\newblock \bibinfo{title}{Incorporating non-local information into information
  extraction systems by gibbs sampling},
\newblock in: \bibinfo{editor}{K.~Knight}, \bibinfo{editor}{H.~T. Ng},
  \bibinfo{editor}{K.~Oflazer} (Eds.), \bibinfo{booktitle}{{ACL} 2005, 43rd
  Annual Meeting of the Association for Computational Linguistics, Proceedings
  of the Conference, 25-30 June 2005, University of Michigan, {USA}},
  \bibinfo{publisher}{The Association for Computer Linguistics},
  \bibinfo{year}{2005}, pp. \bibinfo{pages}{363--370}.
%Type = Article
\bibitem[{Yosef et~al.(2011)Yosef, Hoffart, Bordino, Spaniol, and
  Weikum}]{DBLP:journals/pvldb/YosefHBSW11}
\bibinfo{author}{M.~A. Yosef}, \bibinfo{author}{J.~Hoffart},
  \bibinfo{author}{I.~Bordino}, \bibinfo{author}{M.~Spaniol},
  \bibinfo{author}{G.~Weikum},
\newblock \bibinfo{title}{{AIDA:} an online tool for accurate disambiguation of
  named entities in text and tables},
\newblock \bibinfo{journal}{{PVLDB}} \bibinfo{volume}{4} (\bibinfo{year}{2011})
  \bibinfo{pages}{1450--1453}.
%Type = Inproceedings
\bibitem[{Mendes et~al.(2011)Mendes, Jakob, Garc{\'{\i}}a{-}Silva, and
  Bizer}]{DBLP:conf/i-semantics/MendesJGB11}
\bibinfo{author}{P.~N. Mendes}, \bibinfo{author}{M.~Jakob},
  \bibinfo{author}{A.~Garc{\'{\i}}a{-}Silva}, \bibinfo{author}{C.~Bizer},
\newblock \bibinfo{title}{Dbpedia spotlight: shedding light on the web of
  documents},
\newblock in: \bibinfo{editor}{C.~Ghidini}, \bibinfo{editor}{A.~N. Ngomo},
  \bibinfo{editor}{S.~N. Lindstaedt}, \bibinfo{editor}{T.~Pellegrini} (Eds.),
  \bibinfo{booktitle}{Proceedings the 7th International Conference on Semantic
  Systems, {I-SEMANTICS} 2011, Graz, Austria, September 7-9, 2011}, {ACM}
  International Conference Proceeding Series, \bibinfo{publisher}{{ACM}},
  \bibinfo{year}{2011}, pp. \bibinfo{pages}{1--8}. \URLprefix
  \url{https://doi.org/10.1145/2063518.2063519}.
  \DOIprefix\doi{10.1145/2063518.2063519}.
%Type = Inproceedings
\bibitem[{Usbeck et~al.(2014)Usbeck, Ngomo, R{\"{o}}der, Gerber, Coelho, Auer,
  and Both}]{DBLP:conf/semweb/UsbeckNRGCAB14}
\bibinfo{author}{R.~Usbeck}, \bibinfo{author}{A.~N. Ngomo},
  \bibinfo{author}{M.~R{\"{o}}der}, \bibinfo{author}{D.~Gerber},
  \bibinfo{author}{S.~A. Coelho}, \bibinfo{author}{S.~Auer},
  \bibinfo{author}{A.~Both},
\newblock \bibinfo{title}{{AGDISTIS} - graph-based disambiguation of named
  entities using linked data},
\newblock in: \bibinfo{editor}{P.~Mika}, \bibinfo{editor}{T.~Tudorache},
  \bibinfo{editor}{A.~Bernstein}, \bibinfo{editor}{C.~Welty},
  \bibinfo{editor}{C.~A. Knoblock}, \bibinfo{editor}{D.~Vrandecic},
  \bibinfo{editor}{P.~T. Groth}, \bibinfo{editor}{N.~F. Noy},
  \bibinfo{editor}{K.~Janowicz}, \bibinfo{editor}{C.~A. Goble} (Eds.),
  \bibinfo{booktitle}{The Semantic Web - {ISWC} 2014 - 13th International
  Semantic Web Conference, Riva del Garda, Italy, October 19-23, 2014.
  Proceedings, Part {I}}, volume \bibinfo{volume}{8796} of
  \textit{\bibinfo{series}{Lecture Notes in Computer Science}},
  \bibinfo{publisher}{Springer}, \bibinfo{year}{2014}, pp.
  \bibinfo{pages}{457--471}. \URLprefix
  \url{https://doi.org/10.1007/978-3-319-11964-9\_29}.
  \DOIprefix\doi{10.1007/978-3-319-11964-9\_29}.
%Type = Article
\bibitem[{Moro et~al.(2014)Moro, Raganato, and
  Navigli}]{DBLP:journals/tacl/0001RN14}
\bibinfo{author}{A.~Moro}, \bibinfo{author}{A.~Raganato},
  \bibinfo{author}{R.~Navigli},
\newblock \bibinfo{title}{Entity linking meets word sense disambiguation: a
  unified approach},
\newblock \bibinfo{journal}{Trans. Assoc. Comput. Linguistics}
  \bibinfo{volume}{2} (\bibinfo{year}{2014}) \bibinfo{pages}{231--244}.
%Type = Article
\bibitem[{Egenhofer and Franzosa(1991)}]{DBLP:journals/gis/EgenhoferF91}
\bibinfo{author}{M.~J. Egenhofer}, \bibinfo{author}{R.~D. Franzosa},
\newblock \bibinfo{title}{Point set topological relations},
\newblock \bibinfo{journal}{International Journal of Geographical Information
  Systems} \bibinfo{volume}{5} (\bibinfo{year}{1991})
  \bibinfo{pages}{161--174}. \URLprefix
  \url{https://doi.org/10.1080/02693799108927841}.
  \DOIprefix\doi{10.1080/02693799108927841}.
%Type = Article
\bibitem[{Frank(1992)}]{DBLP:journals/vlc/Frank92}
\bibinfo{author}{A.~U. Frank},
\newblock \bibinfo{title}{Qualitative spatial reasoning about distances and
  directions in geographic space},
\newblock \bibinfo{journal}{J. Vis. Lang. Comput.} \bibinfo{volume}{3}
  (\bibinfo{year}{1992}) \bibinfo{pages}{343--371}. \URLprefix
  \url{https://doi.org/10.1016/1045-926X(92)90007-9}.
  \DOIprefix\doi{10.1016/1045-926X(92)90007-9}.
%Type = Inproceedings
\bibitem[{Skiadopoulos and Koubarakis(2001)}]{DBLP:conf/ssd/SkiadopoulosK01}
\bibinfo{author}{S.~Skiadopoulos}, \bibinfo{author}{M.~Koubarakis},
\newblock \bibinfo{title}{Composing cardinal direction relations},
\newblock in: \bibinfo{editor}{C.~S. Jensen}, \bibinfo{editor}{M.~Schneider},
  \bibinfo{editor}{B.~Seeger}, \bibinfo{editor}{V.~J. Tsotras} (Eds.),
  \bibinfo{booktitle}{Advances in Spatial and Temporal Databases, 7th
  International Symposium, {SSTD} 2001, Redondo Beach, CA, USA, July 12-15,
  2001, Proceedings}, volume \bibinfo{volume}{2121} of
  \textit{\bibinfo{series}{Lecture Notes in Computer Science}},
  \bibinfo{publisher}{Springer}, \bibinfo{year}{2001}, pp.
  \bibinfo{pages}{299--320}.
%Type = Article
\bibitem[{Clementini and Felice(1996)}]{DBLP:journals/isci/ClementiniF96}
\bibinfo{author}{E.~Clementini}, \bibinfo{author}{P.~D. Felice},
\newblock \bibinfo{title}{A model for representing topological relationships
  between complex geometric features in spatial databases},
\newblock \bibinfo{journal}{Inf. Sci.} \bibinfo{volume}{90}
  (\bibinfo{year}{1996}) \bibinfo{pages}{121--136}. \URLprefix
  \url{https://doi.org/10.1016/0020-0255(95)00289-8}.
  \DOIprefix\doi{10.1016/0020-0255(95)00289-8}.
%Type = Article
\bibitem[{van Kreveld and Reinbacher(2004)}]{DBLP:journals/ijcga/KreveldR04}
\bibinfo{author}{M.~J. van Kreveld}, \bibinfo{author}{I.~Reinbacher},
\newblock \bibinfo{title}{Good news: Partitioning a simple polygon by compass
  directions},
\newblock \bibinfo{journal}{Int. J. Comput. Geometry Appl.}
  \bibinfo{volume}{14} (\bibinfo{year}{2004}) \bibinfo{pages}{233--259}.
  \URLprefix \url{https://doi.org/10.1142/S0218195904001469}.
  \DOIprefix\doi{10.1142/S0218195904001469}.
%Type = Inproceedings
\bibitem[{Regalia et~al.(2016)Regalia, Janowicz, and
  Gao}]{DBLP:conf/esws/RegaliaJG16}
\bibinfo{author}{B.~Regalia}, \bibinfo{author}{K.~Janowicz},
  \bibinfo{author}{S.~Gao},
\newblock \bibinfo{title}{{VOLT:} {A} provenance-producing, transparent
  {SPARQL} proxy for the on-demand computation of linked data and its
  application to spatiotemporally dependent data},
\newblock in: \bibinfo{editor}{H.~Sack}, \bibinfo{editor}{E.~Blomqvist},
  \bibinfo{editor}{M.~d'Aquin}, \bibinfo{editor}{C.~Ghidini},
  \bibinfo{editor}{S.~P. Ponzetto}, \bibinfo{editor}{C.~Lange} (Eds.),
  \bibinfo{booktitle}{The Semantic Web. Latest Advances and New Domains - 13th
  International Conference, {ESWC} 2016, Heraklion, Crete, Greece, May 29 -
  June 2, 2016, Proceedings}, volume \bibinfo{volume}{9678} of
  \textit{\bibinfo{series}{Lecture Notes in Computer Science}},
  \bibinfo{publisher}{Springer}, \bibinfo{year}{2016}, pp.
  \bibinfo{pages}{523--538}. \URLprefix
  \url{https://doi.org/10.1007/978-3-319-34129-3\_32}.
  \DOIprefix\doi{10.1007/978-3-319-34129-3\_32}.
%Type = Article
\bibitem[{Shariff et~al.(1998)Shariff, Egenhofer, and
  Mark}]{DBLP:journals/gis/ShariffEM98}
\bibinfo{author}{A.~R. B.~M. Shariff}, \bibinfo{author}{M.~J. Egenhofer},
  \bibinfo{author}{D.~M. Mark},
\newblock \bibinfo{title}{Natural-language spatial relations between linear and
  areal objects: The topology and metric of english-language terms},
\newblock \bibinfo{journal}{International Journal of Geographical Information
  Science} \bibinfo{volume}{12} (\bibinfo{year}{1998})
  \bibinfo{pages}{215--245}.
%Type = Inproceedings
\bibitem[{Mark and Egenhofer(1995)}]{mark1995topology}
\bibinfo{author}{D.~M. Mark}, \bibinfo{author}{M.~J. Egenhofer},
\newblock \bibinfo{title}{Topology of prototypical spatial relations between
  lines and regions in english and spanish},
\newblock in: \bibinfo{booktitle}{AUTOCARTO-CONFERENCE-}, \bibinfo{year}{1995},
  pp. \bibinfo{pages}{245--254}.
%Type = Inproceedings
\bibitem[{Dube and Egenhofer(2012)}]{DBLP:conf/giscience/DubeE12}
\bibinfo{author}{M.~P. Dube}, \bibinfo{author}{M.~J. Egenhofer},
\newblock \bibinfo{title}{An ordering of convex topological relations},
\newblock in: \bibinfo{editor}{N.~Xiao}, \bibinfo{editor}{M.~Kwan},
  \bibinfo{editor}{M.~F. Goodchild}, \bibinfo{editor}{S.~Shekhar} (Eds.),
  \bibinfo{booktitle}{Geographic Information Science - 7th International
  Conference, GIScience 2012, Columbus, OH, USA, September 18-21, 2012.
  Proceedings}, volume \bibinfo{volume}{7478} of
  \textit{\bibinfo{series}{Lecture Notes in Computer Science}},
  \bibinfo{publisher}{Springer}, \bibinfo{year}{2012}, pp.
  \bibinfo{pages}{72--86}. \URLprefix
  \url{https://doi.org/10.1007/978-3-642-33024-7\_6}.
  \DOIprefix\doi{10.1007/978-3-642-33024-7\_6}.
%Type = Inproceedings
\bibitem[{Stocker et~al.(2008)Stocker, Seaborne, Bernstein, Kiefer, and
  Reynolds}]{DBLP:conf/www/StockerSBKR08}
\bibinfo{author}{M.~Stocker}, \bibinfo{author}{A.~Seaborne},
  \bibinfo{author}{A.~Bernstein}, \bibinfo{author}{C.~Kiefer},
  \bibinfo{author}{D.~Reynolds},
\newblock \bibinfo{title}{{SPARQL} basic graph pattern optimization using
  selectivity estimation},
\newblock in: \bibinfo{editor}{J.~Huai}, \bibinfo{editor}{R.~Chen},
  \bibinfo{editor}{H.~Hon}, \bibinfo{editor}{Y.~Liu}, \bibinfo{editor}{W.~Ma},
  \bibinfo{editor}{A.~Tomkins}, \bibinfo{editor}{X.~Zhang} (Eds.),
  \bibinfo{booktitle}{Proceedings of the 17th International Conference on World
  Wide Web, {WWW} 2008, Beijing, China, April 21-25, 2008},
  \bibinfo{publisher}{{ACM}}, \bibinfo{year}{2008}, pp.
  \bibinfo{pages}{595--604}. \URLprefix
  \url{https://doi.org/10.1145/1367497.1367578}.
  \DOIprefix\doi{10.1145/1367497.1367578}.
%Type = Inproceedings
\bibitem[{Jia et~al.(2018)Jia, Abujabal, Roy, Str{\"{o}}tgen, and
  Weikum}]{DBLP:conf/cikm/JiaARSW18}
\bibinfo{author}{Z.~Jia}, \bibinfo{author}{A.~Abujabal}, \bibinfo{author}{R.~S.
  Roy}, \bibinfo{author}{J.~Str{\"{o}}tgen}, \bibinfo{author}{G.~Weikum},
\newblock \bibinfo{title}{{TEQUILA:} temporal question answering over knowledge
  bases},
\newblock in: \bibinfo{editor}{A.~Cuzzocrea}, \bibinfo{editor}{J.~Allan},
  \bibinfo{editor}{N.~W. Paton}, \bibinfo{editor}{D.~Srivastava},
  \bibinfo{editor}{R.~Agrawal}, \bibinfo{editor}{A.~Z. Broder},
  \bibinfo{editor}{M.~J. Zaki}, \bibinfo{editor}{K.~S. Candan},
  \bibinfo{editor}{A.~Labrinidis}, \bibinfo{editor}{A.~Schuster},
  \bibinfo{editor}{H.~Wang} (Eds.), \bibinfo{booktitle}{Proceedings of the 27th
  {ACM} International Conference on Information and Knowledge Management,
  {CIKM} 2018, Torino, Italy, October 22-26, 2018}, \bibinfo{publisher}{{ACM}},
  \bibinfo{year}{2018}, pp. \bibinfo{pages}{1807--1810}.

\end{thebibliography}






% \end{thebibliography}

\end{document}